\documentclass[preprint,onecolumn,nofootinbib,showkeys]{revtex4}
\usepackage{allpurpose}
\usepackage[colorlinks=true,linkcolor=blue,urlcolor=blue,filecolor=black,citecolor=red,pdfstartview=FitV,pdftitle={},pdfsubject={},pdfkeywords={},pdfpagemode=None,bookmarksopen=true]{hyperref}
\usepackage{graphicx}
\usepackage{amsmath}
\usepackage{amsfonts}
\usepackage{amssymb,ulem}
\usepackage{color,xcolor}%
\usepackage{subfigure}
\usepackage{amsthm,amsmath,amssymb}
\usepackage{mathrsfs}
\usepackage{multirow}

\usepackage{dcolumn}
\usepackage{float}
\usepackage{ulem}
\usepackage{color}

\newcommand{\delred}[1]{{\color{red}{\ifmmode\text{\sout{\ensuremath{#1}}}\else\sout{#1}\fi}}}

\usepackage[utf8]{inputenc}

\begin{document}
	
\title{Absorption and scattering of charged scalar waves by charged Horndeski black hole}
	
\author{Qian Li}
\address{School of Physics and Technology, Wuhan University, Wuhan, 430072, China}
\author{Qianchuan Wang}
\address{School of Physics and Technology, Wuhan University, Wuhan, 430072, China}

\author{Junji Jia}
\email[Corresponding author:~]{junjijia@whu.edu.cn}
\address{Department of Astronomy $\&$ MOE Key Laboratory of Artificial Micro- and Nano-structures, School of Physics and Technology, Wuhan University, Wuhan, 430072, China}
	
\begin{abstract}
 We investigate the absorption and scattering of a charged massive scalar field by a charged Horndeski black hole using both the approximation or classical geometric method and the partial wave method and compare the numerical and analytical results, which are found to agree with each other very well.  We observe that an increase in either the BH charge $Q$ or the field charge $q$ when $qQ>0$ leads to a smaller absorption cross section and a widening of the interference fringes in the scattering cross section, while the increase in the field mass enlarges the absorption cross section and the width of the interference fringes. Compared to the Reissner-Nordstr$\ddot{\rm{o}}$m BH with the same charge and other parameter settings, the absorption and scattering cross sections of the charged Horndeski BH are higher, and its interference fringes are narrower. We also investigate the effect of the field charge $q$ on the absorption and scattering cross sections when superradiance is triggered. It is shown that the total absorption cross section can be negative, and the scattering intensity can be significantly enhanced by superradiance.
\end{abstract}


\maketitle
	
\section{introduction}\label{introduction}

	At galactic and cosmological scales, general relativity (GR) can successfully explain many gravitational phenomena, including but not limited to the prediction and interpretation of planetary orbits, gravitational lensing, and the formation of black holes (BHs).  However, despite its successes, GR may not fully explain some phenomena, such as the accelerated expansion of the universe without the help of the yet-to-be-discovered dark matter, and dark energy,  prompting the exploration of alternative theories.  Among the modified theories of gravity, scalar-tensor theories, which involve the non-minimal couplings of a scalar field and a metric tensor, are considered the most natural and simplest modification of GR. In the 1970s, Horndeski proposed the most general scalar-tensor theory, which possesses field equations of the second order and an energy-momentum tensor of the second order \cite{Horndeski:1974wa}. There are many fascinating characteristics of Horndeski gravity, the most striking of which is the non-minimal derivative coupling composed of the scalar field and the Einstein tensor.  This term plays an important role in explaining the accelerated expansion in the absence of any scalar potential \cite{Amendola:1993uh}. Modified gravities that belong to this class have received exhaustive treatment in astrophysics and cosmology \cite{Saridakis:2010mf,Charmousis:2011bf,Maselli:2016gxk,Kobayashi:2019hrl, Galeev:2021xit}. Besides, there has also been a great deal of attention paid to BH solutions in the context of Horndeski theory \cite{Rinaldi:2012vy,Anabalon:2013oea,Cisterna:2014nua,Maselli:2015yva,Babichev:2016fbg,Antoniou:2017hxj,Babichev:2017guv}.

Scattering phenomena play a crucial role in physics, not only in understanding the atomic nucleus but also in exploring the nature and behavior of BHs. Since the 1960s, considerable work has been carried out on the absorption and scattering of BHs for waves with different spins $s$, namely, massless and massive scalar ($s=0$) \cite{Sanchez:1977si,Sanchez:1977vz,Crispino:2007zz,Crispino:2009ki,Chen:2011jgd,Macedo:2013afa,Macedo:2014uga,Anacleto:2019tdj,Lima:2020auu,Anacleto:2020lel,Li:2021epb,Li:2022wzi,Sun:2023woa,Wan:2022vcp,Jung:2004yh,Benone:2014qaa,deCesare:2023rmg}, Dirac ($s=1/2$) \cite{Dolan:2006vj,Sporea:2017zxe}, electromagnetic ($s=1$) \cite{Crispino:2007qw,Crispino:2008zz,Crispino:2009xt,Leite:2017zyb,Leite:2018mon,deOliveira:2019tlk}, and gravitational  ($s=2$)  \cite{Dolan:2008kf,Crispino:2015gua} waves. During this period, some scattering properties were discovered, such as glory \cite{Matzner:1985rjn}, orbiting \cite{Anninos:1992ih}, and superradiant scattering. The first two features, related to the classical deflection angle, are relatively straightforward \cite{Ford:2000uye}. For superradiant scattering \cite{Misner:1972kx}, when a boson field (with the spin $s$ being an integer) impinges upon a BH, there is a possibility that the energy of  BH  will transfer to the scattered waves. In other words, similar to the Penrose process, scattered waves can extract energy from BHs. It is worth mentioning that, apart from rotating BHs, the energy of charged BHs can be extracted by a charged massive scalar field \cite{Brito:2015oca}. More specifically, superradiant scattering can lead to a negative absorption cross  section \cite{Benone:2015bst,Benone:2019all,dePaula:2024xnd} and may enhance the scattering intensity in such kind of scattering in a certain range of wave frequency when certain conditions are met. However, it is also known that when the scalar field hits a Kerr BH, superradiance has a negligible effect on the scattered flux \cite{Glampedakis:2001cx}. Therefore, in this work we aim to investigate whether the superradiance can also occur in charged Horndeski BHs, and if so, whether it is also suppressed.

In this paper, we focus on a charged Horndeski BH solution where the derivative of a scalar field is coupled to the Einstein tensor in the presence of an electric field. Previously, the thermodynamic properties \cite{Feng:2015wvb}, the weak and strong deflection gravitational lensings \cite{Wang:2019cuf}, and the shadow images and rings \cite{Gao:2023mjb} of this charged BH have been studied by different authors. However, a study of the scattering problem of this BH is still lacking, and this will be the main purpose of this work.
	
This paper is organized as follows: In Sec. \ref{spacetime}, we briefly introduce the charged Horndeski BH solution, which is a specific case of the Horndeski theory.  The absorption and scattering cross sections are obtained using the partial wave analysis in Sec. \ref{numerical}. Section \ref{analyical} is dedicated to studying the cross section in the low- and high-frequency limits.  We present and analyze our numerical results of the absorption and scattering cross sections in Sec. \ref{results}. The influences of superradiance on absorption and scattering are discussed in Sec. \ref{superradiance}. Finally, we draw conclusions from our results in Sec. \ref{conclusion}. For consistency, we use the metric signature $(-+++)$ and natural units ($c=\hbar=G= 4\pi\varepsilon_0=1$). 
	
\section{charged Horndeski BH}\label{spacetime}
	
In this section, we describe a charged Horndeski BH solution given in Ref. \cite{Cisterna:2014nua}. This solution originates from a specific model, whose action consists of the Einstein-Hilbert term, the non-minimal kinetic term of the scalar field $\varphi$ coupled with the Einstein tensor with coupling strength $\eta$, and the Maxwell term \cite{Cisterna:2014nua} 
	\begin{align}\label{action}
		S=\int \dd^4x\sqrt{-g}\left[\dfrac{R}{16\pi}+\dfrac{\eta}{2}G_{\mu\nu}\nabla^\mu\varphi\nabla^\nu\varphi-\dfrac{1}{4}F_{\mu\nu}F^{\mu\nu}\right].
	\end{align}
The theory has a charged Horndeski BH solution, which is asymptotically Minkowski and described by the line element 
	\begin{align} \label{lm}
		\dd s^2 = - A(r) \dd t^2 +B(r) \dd r^2 + r^2 \left(\dd\theta^2 + \sin^2 \theta \dd\phi^2\right),
	\end{align}
	with 
		\begin{align}\label{A,B}
		A(r)&=1-\dfrac{2M}{r}+\dfrac{Q^2}{4r^2}-\dfrac{Q^4}{192r^4},\\
	 B(r)&=\left(1-\frac{Q^2}{8r^2}\right)^2 A^{-1}(r),
	\end{align}
where $M$ is the BH mass and  $Q$ stands for the electric charge. This solution degenerates to the Schwarzschild BH when $Q=0$, but can't recover to the Reissner-Nordstr$\ddot{\rm{o}}$m (RN) BH. Its event horizon $r_h$ is the largest real root of $A(r)=0$. By analyzing the event horizon and curvature singularities, which lie within the horizon, the charge needs to satisfy the condition \cite{Feng:2015wvb}
	\begin{align}\label{condition}
		|Q|<\frac{3M}{\sqrt{2}}.
\end{align}	
For future reference, here  we also list the metric of the RN BH
 \begin{align}\label{A,BRN}
		A_{\rm{RN}}(r)=B_{\rm{RN}}^{-1}(r)=1-\dfrac{2M}{r}+\dfrac{Q^2}{r^2}, 
	\end{align}
where the value of the charge satisfies the condition $|Q|<M$.

\section{partial wave approach}  \label{numerical}
For a charged massive scalar field  propagating in a static BH, the Klein-Gordon equation for this scalar field in this BH spacetime can be described as
\begin{align}	\label{kge}
	\left(\nabla_{\nu} - i q A_{\nu}\right)\left(\nabla^{\nu} - i q A^{\nu}\right) \Phi_{\omega l}-m^2  \Phi_{\omega l}=0,
\end{align}
where  $m$ and $q$ are the mass and electric charge of the scalar field with frequency $\omega$, respectively. Besides,  $A_{\mu}$ is   the electromagnetic four-potential, given by
\begin{align} 
	A_{0}(r)= -\frac{Q}{r}+\frac{Q^3}{24r^3}  \label{eq:A0}
\end{align}
 and the remaining components are zero. In comparison, for the RN metric, the non-zero four-potential is $A^{\rm{RN}}_{0}(r)= -\frac{Q}{r}$. Since we only focus on the scattering properties of the field in this work, the scalar wave will be able to propagate to infinity, and therefore the frequency will satisfy $\omega > m$. 
This equation can be solved using a separation of variables ansatz, namely, 
\begin{align}	\label{eq:separation}
	\Phi_{\omega l} = \frac{\psi_{\omega l}(r)}{r} P_{l}(\cos\theta) \expe^{-i\omega t},~(l=0,1,\cdots),
\end{align}
where $P_{l}(\cos\theta)$ are the Legendre polynomials. It is easy to find that the radial function $\psi_{\omega l}(r)$ satisfies the equation
\begin{align}	\label{fdr}
	\sqrt{\frac{A(r)}{B(r)}}\frac{\dd}{\dd r}\left(\sqrt{\frac{A(r)}{B(r)}}\frac{\dd}{\dd r}\psi_{\omega l}\right) -V_\textit{eff}(r)\psi_{\omega l}=0,
\end{align}
with the effective potential
\begin{align}\label{veff}
	V_\textit{eff}(r) =-\big(\omega+q A_0(r)\big)^2 +A(r)\left[m^2+\frac{l(l+1)}{r^2}+\frac{1}{2r A(r) }\frac{\dd}{\dd r}\frac{A(r)}{B(r)}\right].
\end{align}	
Defining the tortoise coordinate $r_*$ to satisfy
\begin{align}\label{tortoise}
	\frac{\dd r} { dr_*} =\sqrt{\frac{A(r)}{B(r)}},
\end{align}
we get from Eq. \eqref{fdr} a Schr$\ddot{\textrm{o}}$dinger-like equation
\begin{align}\label{Scheq}
	\frac{\dd ^2}{\dd r_*^2}\psi_{\omega l}-V_\textit{eff}(r) \psi_{\omega l} = 0.	
\end{align}

The effective potential has a local maximum in the intermediate region of $r$ and tends to $-k_h^2$  and $-k_\infty^2$ at the event horizon and spatial infinity, respectively. The asymptotic behavior of the solution of Eq. \eqref{Scheq} in the scattering problem is described by 
\begin{align}\label{solution}
	\psi_{\omega l}(r) \approx
	\left\{
	\begin{array}{ll}
		\expe^{-ik_\infty r_*} + R_{\omega l} \expe^{ik_\infty r_*}, \quad &\mbox{for $r_*\rightarrow +\infty~ (r\rightarrow \infty)$},\\
		T_{\omega l} \, \expe^{-i k_h r_*}, \quad &\mbox{for $r_*\rightarrow -\infty~(r\rightarrow r_h)$},
	\end{array}
	\right.
\end{align}
where 
\begin{align} k_\infty \equiv \sqrt{\omega^2-m^2}, ~ k_h \equiv \omega + q A_0(r_h)
\end{align}
and $R_{\omega l}$ and $T_{\omega l}$ are the reflection and transmission coefficients, respectively. To obtain the relationship between $R_{\omega l} $ and $T_{\omega l} $, we calculate  the Wronskian at the event horizon and infinity
\begin{align}
	W_h&=\left(\psi^h_{\omega l}(r)\frac{\dd \psi^{* h}_{\omega l}(r)}{\dd r_*}-\psi^{* h}_{\omega l}(r) \frac{\dd \psi^{h}_{\omega l}(r)}{\dd r_*}\right),\\
	W_\infty&=\left(\psi^\infty_{\omega l}(r)\frac{\dd \psi^{* \infty}_{\omega l}(r)}{\dd r_*}-\psi^{* \infty}_{\omega l}(r) \frac{\dd \psi^{\infty}_{\omega l}(r)}{\dd r_*}\right) .
\end{align}
Then, using the conservation of the Wronskian, the relationship between $R_{\omega l}$ and $T_{\omega l}$ can be expressed by   
\begin{align}\label{rtf}
	\left|R_{\omega l}\right|^2 + \frac{k_h}{k_\infty} \left|T_{\omega l}\right|^2 = 1.
\end{align}

In the partial wave approach, the total absorption cross section for a scalar wave impinging upon a spherically symmetric BH is given by \cite{Sanchez:1977si} 
\begin{align} \label{tacs}
	\sigma_{\mathrm{abs}} =\sum_{l=0}^{\infty}\sigma_l,
\end{align}
where $\sigma_l$ is the partial absorption cross section and can be expressed as
\begin{align} \label{pacs}
	\sigma_l= \frac{\pi}{k_\infty^2}(2l+1)\left(1-\left|R_{\omega l} \right|^2 \right).
\end{align}

And the total differential scattering cross section is given by
\begin{align} \label{dscs}
	\frac{\dd \sigma}{\dd \Omega} =\left|f(\theta)\right|^2,
\end{align}
where scattering amplitude $f(\theta)$ can be expressed as \cite{Sanchez:1977vz}:
\begin{align}\label{sa}
	f(\theta) =\frac{1}{2ik_\infty}\sum_{l=0}^{\infty}(2l+1)\left[(-1)^{l+1}R_{\omega l}-1\right]	P_{l}(\cos\theta).
\end{align}
  We will use Eqs. \eqref{tacs} and \eqref{sa} to numerically study these cross sections.

\section{analyical results}\label{analyical}

\subsection{Low-frequency limit}\label{analyical-low}
 In this subsection, we obtain the approximate analytical expressions for the absorption cross section in the low-frequency limit. Similar to the calculation process described in Ref. \cite{Unruh:1976fm}, we will calculate the solution of the Schr$\ddot{\textrm{o}}$dinger-like Eq. \eqref{Scheq} in three regions: (i) the region approaching the BH event horizon ($r\rightarrow r_h$); (ii) considering  $M\omega\ll1$, $Mm\ll1$ and  $Mq\ll 1$
in the intermediate region; (iii) the region far from the BH.  Then, we will match local solutions between the three regions to obtain the reflection coefficient $R_{\omega l}$.

 Firstly, by substituting $\psi_{\omega l}$ with $\psi r$, we can reformulate the radial   Eq. \eqref{Scheq}  as
 \begin{align}	\label{fdr1}
 \frac{1}{r^2}	\sqrt{\frac{A(r)}{B(r)}}\frac{\dd}{\dd r}\left( r^2\sqrt{\frac{A(r)}{B(r)}}\frac{\dd}{\dd r}\psi\right) +\left[\big(\omega+q A_0(r)\big)^2 -A(r)\left(m^2+\frac{l(l+1)}{r^2}\right)\right]\psi=0.
 \end{align}
 Besides, for later calculations, $A(r)$ can also be rewritten by
  \begin{align}	\label{A1}
 	A(r)&=\frac{(r-r_{1})(r-r_2)(r-r_3)(r-r_4)}{r^4}, 
 \end{align}
 where $r_{i}$ (for $i=1,2,3,4$) is the root of $A(r)=0$ and $r_{4}=r_{h}$.
 
For region (I) ($r \approx r_{4}$), the Eq. \eqref{fdr1} can be approximated as 
 \begin{align}	\label{region1}
 	\frac{\dd ^2}{\dd r_*^2}\psi^{I}+ k_{h}^2 \psi^{I}= 0,	
 \end{align}
and   $\psi^{I}\propto\expe^{-ik_{h}r_*}$. By utilizing Eqs. \eqref{tortoise} and \eqref{A1}, $r_*$ can be expressed as a function of $r$, specifically
  \begin{align}	\label{tortoise1}
r_*&=r-r_{1}
-
\frac{r_{1}^2\left(Q^2-8r_{1}^2\right)\ln{\left(r- r_{1}\right)}}{8{\left(r_{1}- r_{2}\right)}{\left(r_{1}- r_{3}\right)}{\left(r_{1}- r_{4}\right)}}
-
\frac{r_{2}^2\left(Q^2-8r_{2}^2\right)\ln{\left(r- r_{2}\right)}}{8{\left(r_{2}- r_{1}\right)}{\left(r_{2}- r_{3}\right)}{\left(r_{2}- r_{4}\right)}} \\ \nonumber
 &-\frac{r_{3}^2\left(Q^2-8r_{3}^2\right)\ln{\left(r- r_{3}\right)}}{8{\left(r_{3}- r_{1}\right)}{\left(r_{3}- r_{2}\right)}{\left(r_{3}- r_{4}\right)}} 
-
\frac{r_{4}^2\left(Q^2-8r_{4}^2\right)\ln{\left(r- r_{4}\right)}}{8{\left(r_{4}- r_{1}\right)}{\left(r_{4}-r_{2}\right)}{\left(r_{4}- r_{3}\right)}}.
 \end{align}
 When $r \approx r_{4}$ and only the main term in the above expression is considered, we have 
   \begin{align}	\label{tortoise2}
 	r_* \sim - \frac{r_{4}^2\left(Q^2-8r_{4}^2\right)\ln{\left(r- r_{4}\right)}}{8{\left(r_{4}- r_{1}\right)}{\left(r_{4}-r_{2}\right)}{\left(r_{4}- r_{3}\right)}}+ r_*^c,
 \end{align}
 where $r_*^c$ is a constant. Therefore, the solution of  the region (I) can be given as
 \begin{align}	\label{solution1}
\psi^{I}= T_{\omega 0} \left|r-r_{4} \right|^{-ik_{h}\alpha},
\end{align}
where  
 \begin{align}	\label{alpha}
\alpha=-
\frac{r_{4}^2\left(Q^2-8r_{4}^2\right)}{8{\left(r_{4}- r_{1}\right)}{\left(r_{4}-r_{2}\right)}{\left(r_{4}- r_{3}\right)}}.
\end{align}

For region (II), we consider the limit $M\omega\ll1$, $Mm\ll1$ and  $Mq\ll 1$ in Eq. \eqref{fdr1}. Since the $l=0$ mode contributes the most to the absorption cross section in this limit \cite{Unruh:1976fm}, we focus only on this mode. The radial Eq. \eqref{fdr1} for this region reduces to 
 \begin{align}	\label{region2}
	\frac{\dd ^2 \psi^{II}}{\dd r^2}+ \frac{1}{2}\left(\frac{4}{r}+ \frac{A^\prime(r)}{A(r)}-\frac{B^\prime(r)}{B(r)} \right) \frac{\dd \psi^{II}}{\dd r}= 0,	
\end{align}
and its solution is given by 
 \begin{align}	\label{solution2}
	\psi^{II}= a_2-a_1 \bigg[&
	\frac{\left(Q^2-8r_{1}^2\right)\ln{\left(r- r_{1}\right)}}{{\left(r_{1}- r_{2}\right)}{\left(r_{1}- r_{3}\right)}{\left(r_{1}- r_{4}\right)}}
    +
	\frac{\left(Q^2-8r_{2}^2\right)\ln{\left(r- r_{2}\right)}}{{\left(r_{2}- r_{1}\right)}{\left(r_{2}- r_{3}\right)}{\left(r_{2}- r_{4}\right)}} \\ \nonumber
	& +\frac{\left(Q^2-8r_{3}^2\right)\ln{\left(r- r_{3}\right)}}{{\left(r_{3}- r_{1}\right)}{\left(r_{3}- r_{2}\right)}{\left(r_{3}- r_{4}\right)}} 
	+
	\frac{\left(Q^2-8r_{4}^2\right)\ln{\left(r- r_{4}\right)}}{{\left(r_{4}- r_{1}\right)}{\left(r_{4}-r_{2}\right)}{\left(r_{4}- r_{3}\right)}}\bigg],
\end{align}
where $a_{1}$ and $a_{2}$ are  integration constants that need to be determined.

In the limit $r \rightarrow r_{4}$,  Eq. \eqref{solution2} reduces to 
 \begin{align}	\label{solution2_2}
	\psi^{II}= a_2-a_1 \bigg[&
	\frac{\left(Q^2-8r_{1}^2\right)\ln{\left(r_{4}- r_{1}\right)}}{{\left(r_{1}- r_{2}\right)}{\left(r_{1}- r_{3}\right)}{\left(r_{1}- r_{4}\right)}}
	+
	\frac{\left(Q^2-8r_{2}^2\right)\ln{\left(r_{4}- r_{2}\right)}}{{\left(r_{2}- r_{1}\right)}{\left(r_{2}- r_{3}\right)}{\left(r_{2}- r_{4}\right)}} \\ \nonumber
	& +\frac{\left(Q^2-8r_{3}^2\right)\ln{\left(r_{4}- r_{3}\right)}}{{\left(r_{3}- r_{1}\right)}{\left(r_{3}- r_{2}\right)}{\left(r_{3}- r_{4}\right)}} 
	+
	\frac{\left(Q^2-8r_{4}^2\right)\ln{\left(r- r_{4}\right)}}{{\left(r_{4}- r_{1}\right)}{\left(r_{4}-r_{2}\right)}{\left(r_{4}- r_{3}\right)}}\bigg],
\end{align}
 and Eq. \eqref{solution1} is rewritten as 
   \begin{align}	\label{solution1_1}
  	\psi^{I}= T_{\omega 0}\left[1-ik_{h} \alpha \ln{\left(r- r_{4}\right)} \right].
  \end{align}
Comparing  the two expressions above, we can obtain the integration constants
   \begin{align} \label{a12}
	a_{1}=& \frac{i T_{\omega 0}  \alpha k_{h} {\left(r_{4}- r_{1}\right)}{\left(r_{4}-r_{2}\right)}{\left(r_{4}- r_{3}\right)}}{Q^2-8r_{4}^2}, \\ \nonumber 
a_{2}=  	&T_{\omega 0}\left(1 -i \alpha k_{h} \beta \right),
\end{align}
with
   \begin{align} \label{beta}
	\beta= &\frac{1 }{{\left(r_{1}- r_{2}\right)}{\left(r_{1}-r_{3}\right)}{\left(r_{2}- r_{3}\right)}\left( Q^2-8r_{4}^2\right)} \\ \nonumber
	& \times \bigg[ \left(Q^2-8r_{1}^2\right) {\left(r_{2}-r_{3}\right)}  {\left(r_{2}-r_{4}\right)}{\left(r_{3}-r_{4}\right)} \ln{\left(r_{4}- r_{1}\right)} \\ \nonumber
	&+  \left(Q^2-8r_{2}^2\right) {\left(r_{1}-r_{3}\right)} {\left(r_{4}-r_{1}\right)}{\left(r_{3}-r_{4}\right)}\ln{\left(r_{4}- r_{2}\right)} \\ \nonumber
	&+ \left(Q^2-8r_{3}^2\right) {\left(r_{2}-r_{1}\right)} {\left(r_{4}-r_{1}\right)}{\left(r_{2}-r_{4}\right)}\ln{\left(r_{4}- r_{3}\right)}
	\bigg].
\end{align}

For region (III) ($r \rightarrow \infty$), we neglect terms of order  $\mathcal{O}(1/r^2)$; consequently, the radial Eq. \eqref{fdr1} reduces to 
  \begin{align}	\label{region3}
 \left[	\frac{\dd ^2}{\dd r^2}+ \left( \omega^2-m^2\right)+\frac{2\left(2M\omega^2- M m^2- qQ\omega \right)}{r}\right]
\left[r \left(\frac{B(r)}{A(r)}\right)^{\frac{1}{4}}	\psi^{III}\right]= 0.	
 \end{align}
When $r \left(\frac{B(r)}{A(r)}\right)^{\frac{1}{4}}\psi^{III}$ is considered as a  new radial function, the solution of Eq. \eqref{region3} is
 \begin{align}	\label{solution3}
	\psi^{III}= c_1 \frac{F_{0}(\eta, \omega v r)}{r}+ c_2 \frac{G_{0}(\eta, \omega v r)} {r}
\end{align}
 where $F_{0}(\eta, \omega v r)$ and $G_{0}(\eta, \omega v r)$ represent the regular and irregular Coulomb wave functions \cite{Abramowitz}, respectively, and
 \begin{align}	\label{eta}
\eta=-\frac{M\left(2\omega^2-m^2 \right)}{\omega v }+\frac{qQ}{v}= -\frac{Mm\left(1+v^2 \right)}{v \sqrt{1-v^2} }+\frac{qQ}{v}
\end{align}
with $v=\sqrt{1-m^2/\omega^2}$ and $\omega v= \sqrt{\omega^2-m^2}=k_{\infty}$. In the far field, the asymptotic expansion of Eq. \eqref{solution3} is 
 \begin{align}\label{solution3_1}
  \psi^{III}=R_{\omega 0}^{ref}\expe^{i \mathcal{\vartheta}_0}+R_{\omega 0}^{inc}\expe^{-i \mathcal{\vartheta}_0},
 \end{align}
where $\mathcal{\vartheta}_0= \omega v r-\eta \ln{(2 \omega v r)}+\arg{\Gamma{(1+i\eta)}}$.   From Eqs. \eqref{solution3} and \eqref{solution3_1}, we have the relationship
 \begin{align}\label{AA}
 R_{\omega 0}^{ref}=\frac{c_{1}+i c_{2}}{2i}, ~ R_{\omega 0}^{inc}=\frac{-c_{1}+i c_{2}}{2i}.
\end{align}

In order to match $\psi^{III}$ with $\psi^{II}$, we consider the limit $\omega v r\ll1$ in Eq. \eqref{solution3}, then  this solution can be simplified to
  \begin{align}\label{solution3_2}
 	\psi^{III}= c_{1}  \rho \omega v + \frac{c_{2}}{\rho r},
 \end{align}
with $\rho^2=\frac{2\pi \eta}{\expe^{2\pi\eta}-1}$. Furthermore, in the far field, we find that the solution \eqref{solution2} can be expanded to
\begin{align}\label{solution2_3}
 \psi^{II}=a_{2}-\frac{8 a_{1}}{r}.
\end{align}
Substituting Eq. \eqref{a12} into Eq. \eqref{solution2_3} and then comparing the result with Eq. \eqref{solution3_2},  $c_{1}$ and $c_{2}$ take the following forms, respectively
\begin{align}\label{c12}
c_{1}=&\frac{T_{\omega 0}\left(1-i\alpha \beta k_{h} \right)}{\omega v \rho }, \\ \nonumber
c_{2}=&-\frac{ 8 i\alpha \rho k_{h} T_{\omega 0}{\left(r_{4}-r_{1}\right)} {\left(r_{4}-r_{2}\right)}{\left(r_{4}-r_{3}\right)}}{Q^2-8r_{4}^2}.
\end{align}
 Insert Eq. \eqref{c12} into Eq. \eqref{AA}, we have 
\begin{align}\label{AA1}
R_{\omega 0}^{ref}&=-\frac{1}{2}iT_{\omega 0}\left[\frac{1}{\omega v \rho}-\frac{i \alpha \beta k_{h} }{\omega v \rho}-
\frac{8 \alpha \rho k_{h}{\left(r_{1}-r_{4}\right)} {\left(r_{4}-r_{2}\right)}{\left(r_{4}-r_{3}\right)} }{Q^2-8r_{4}^2}
\right],\\ \nonumber
R_{\omega 0}^{inc}&=\frac{1}{2}iT_{\omega 0}\left[\frac{1}{\omega v \rho}-\frac{i \alpha \beta k_{h} }{\omega v \rho}+
\frac{8 \alpha \rho k_{h}{\left(r_{1}-r_{4}\right)} {\left(r_{4}-r_{2}\right)}{\left(r_{4}-r_{3}\right)} }{Q^2-8r_{4}^2}
\right].
\end{align}
The reflection coefficient $R_{\omega l}$ is defined as $ R_{\omega 0}^{ref}/R_{\omega 0}^{inc} $, and is   carried into the partial absorption cross section \eqref{pacs} for $l=0$. Finally,  the partial absorption cross section in the  low-frequency limit is given by
\begin{align}\label{low_abs}
 \sigma_{lf}\approx\frac{\pi}{\omega^2 v^2}\left\{\frac{4 r_{4}^2\rho^2 \omega \left(\omega+qA_{0}(r_{4})\right)v }{\left[1+ r_{4}^2\rho^2 \omega \left(\omega+qA_{0}(r_{4})\right)v\right]^2+ \left( \alpha \beta k_{h}\right)^2}
  \right\}.
\end{align}
Considering only the first term in the denominator of the above equation, this cross section can be simplified to  
\begin{align}\label{low_abs_1}
	\sigma_{lf}\approx\frac{\pi}{\omega v}\left[4 r_{4}^2\rho^2  \left(\omega+qA_{0}(r_{4})\right)\right].	 
\end{align}
Using Eq. \eqref{eta}, the  expression for $\rho$ is
 \begin{align}\label{rho}
  \rho^2=\frac{2\pi \eta}{\expe^{2\pi\eta}-1}= 2\pi\left[\frac{Mm\left(1+v^2 \right)}{v \sqrt{1-v^2} }-\frac{qQ}{v} \right]/ \left\{ 1-\exp{\left[-\frac{2\pi Mm\left(1+v^2 \right)}{v \sqrt{1-v^2} }+\frac{2\pi qQ}{v}\right]}\right\},
 \end{align}
which is related to the velocity of this test field. Thus, we define the critical velocity $v_{c}=2\pi\left( M m-qQ\right)$.

For $v> v_{c}$, $\rho^2\sim 1$, this absorption cross section becomes
\begin{align}\label{low_abs1}
\sigma_{lf}^{(1)} = \frac{4 \pi r_{4}^2}{\omega v}(\omega+qA_{0}(r_{h})).
\end{align}
For low velocity $v< v_{c}$, $\rho^2\sim 2\pi\left( M m-qQ\right)/v$, we have 
 \begin{align}\label{low_abs2}
 \sigma_{lf}^{(2)} = \frac{8\pi^2 r_{4}^{2}}{\omega v^2}(\omega+qA_{0}(r_{h}))(m M-q Q). 
\end{align}
If $q=Q=0$, the results (97) and (99) of a Schwarzschild BH in Ref. \cite{Unruh:1976fm} are obtained. 

\subsection{High-frequency limit }\label{analysis-high}
In this subsection, we study the propagation of test particles in the charged Horndeski spacetime to better understand the absorption and scattering cross sections in the high-frequency region.

\subsubsection{Geodesic scattering} \label{analysis_geodesic}
We investigate the geodesic scattering of charged massive particles in a charged Horndeski BH.  Since the charged particle is subject to the Lorentz force, the classical Lagrangian $\mathcal{L}$ is described by \cite{dePaula:2024xnd}
\begin{align}\label{Lag}
			\mathcal{L}=\frac{1}{2} g_{\mu\nu}\dot{x}^\mu\dot{x}^\nu+\frac{q}{m}A_{\mu}\dot{x}^\mu,
\end{align}	
where $\dot{x}^\mu=\dd x^{\mu}/\dd \tau$ with  $\tau$ representing the proper time. For massive particles, we have the timelike normalization condition $g_{\mu\nu}\dot{x}^{\mu}\dot{x}^{\nu}=-1$. Using Eq.  \eqref{Lag} and the above condition, one finds the following equations of motion
	\begin{align} 
		\dot{t}&=\frac{E+qA_0(r)}{mA(r)},\label{t}\\ 
		\dot{\phi}&=\frac{L}{m r^2}, \label{phi}\\ 
		\dot{r}^2&=\frac{\left[\left(E+qA_0(r)\right)^2-m^2A(r)\right]	r^2-L^2A(r)}{m^{2}A(r) B(r)r^2} \label{radial},
	\end{align}
where $E$ and $L$ are two conserved quantities representing the energy and orbital angular momentum of the particle, respectively, and without loss of generality, we have considered a trajectory on the equatorial plane, i.e. $\theta=\pi/2$.  

For the convenience of future calculations, we define the impact parameter as $b=L/vE$ where $v=\sqrt{1-m^2/E^2}$ is the asymptotic velocity  and rewrite the radial equation ($\ref{radial}$) as 
	\begin{eqnarray}
    		\dot{r}^2\big(\frac{m^2}{L^2}\big)=\frac{\big(m+q\sqrt{1-v^2}A_0(r)\big)^2}{A(r) B(r)b^2 v^2 m^2}-\frac{1}{B(r)}\big(\frac{1-v^2}{v^2 b^2}+\frac{1}{r^2}\big).\label{radial1}
	\end{eqnarray}
When $\dot{r}=0$, the trajectory will turn around in the radial direction, whose minimal radial coordinate can be denoted as $r_0$. When $\ddot{r}=0$, then the $r_0$ will reach its critical value $r_c$, below which the trajectory will enter and then be trapped by the BH. Therefore letting Eq. \eqref{radial1} and its first derivative with respect to the proper time or equivalent with respect to $r$ equal zero, we can obtain a set of two equations, which determine the $r_c$ and the corresponding critical impact parameter $b_c$, 
\begin{align}
	&2A(r_c)\left\{m^2 (1-v^2) A(r_c)-\left(m +q \sqrt{1-v^2} A_0(r_c)\right)\left[m +q \sqrt{1-v^2}\big(A_0(r_c)+r_c A^\prime_0(r_c)\big) \right] \right\} \nonumber  \\
	&+r_c A^\prime(r_c) \left(m+q \sqrt{1-v^2} A_0(r_c)\right)^2 =0, 	\label{rc} \\
	&b^2_c=\frac{r^2_c \left(1-v^2\right) \left[\left(\frac{m}{\sqrt{1-v^2}}+q A_0(r_c)\right)^2-m^2 A(r_c)\right]}{v^2 m^2 A(r_c)},
\label{bc}
\end{align}
which are consistent with Eqs. (42) and (43) in Ref. \cite{dePaula:2024xnd}.

\begin{figure}[htp]
	\centering
	\begin{tabular}{cc}
		\includegraphics[width=0.5 \textwidth]{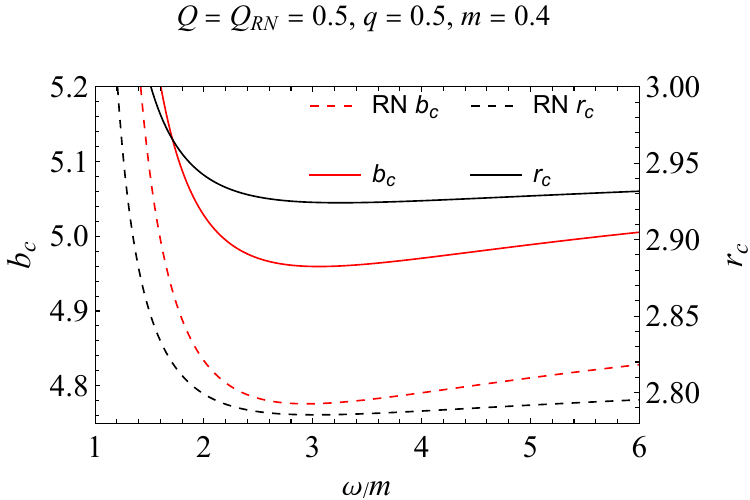}
  	\includegraphics[width=0.5 \textwidth]{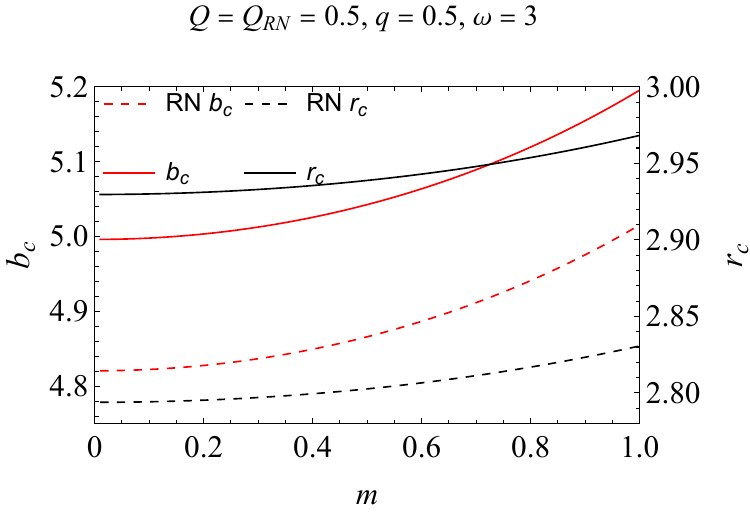}
	\end{tabular}
\caption{	
$r_c$ and $b_c$ of a charged particle with $q=0.5$ as functions of $\omega$ (left) and $m$ (right) in charged Horndeski (solid lines) and RN (dashed lines) BHs with $Q=0.5$.}	\label{rc-bc}	
	\end{figure}
 
\begin{figure}[htp]
		\centering
		\begin{tabular}{cc}
			\includegraphics[width=0.5 \textwidth]{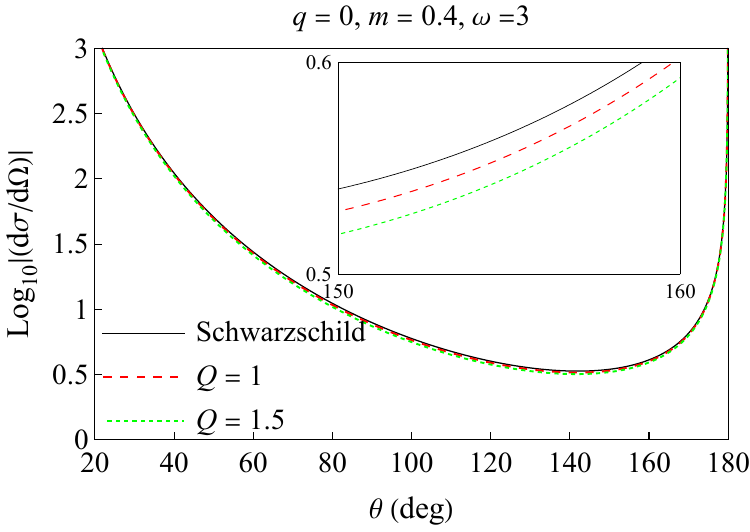}
			\includegraphics[width=0.5 \textwidth]{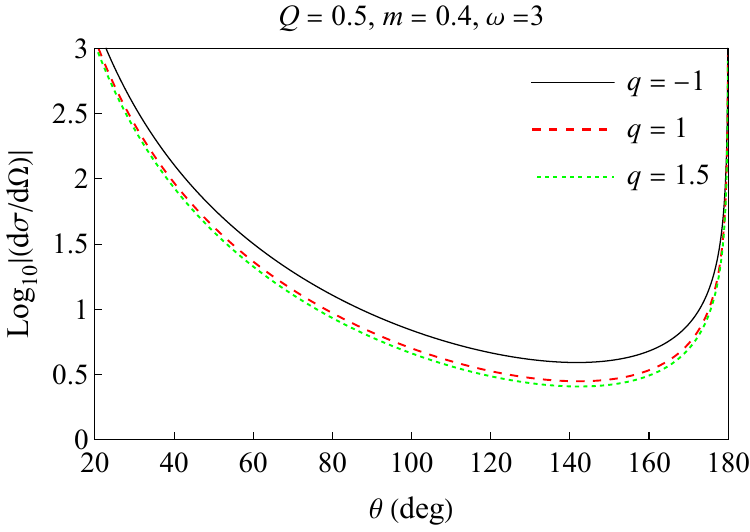}
		\end{tabular}
			\begin{tabular}{cc}
			\includegraphics[width=0.5 \textwidth]{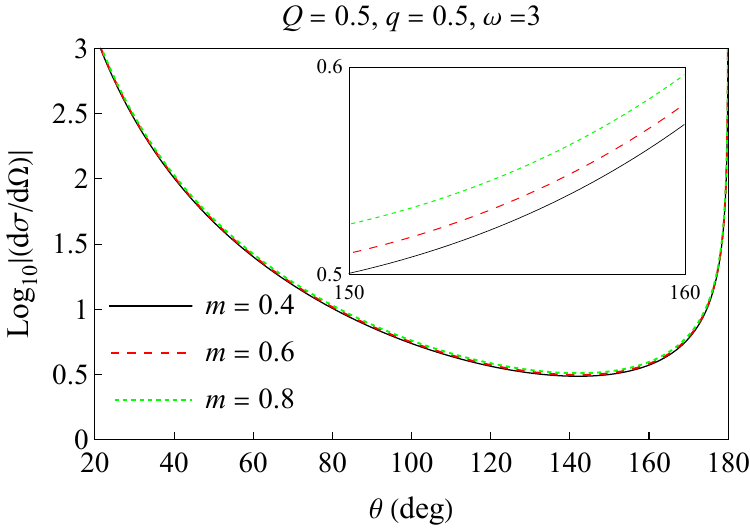}		
			\includegraphics[width=0.5 \textwidth]{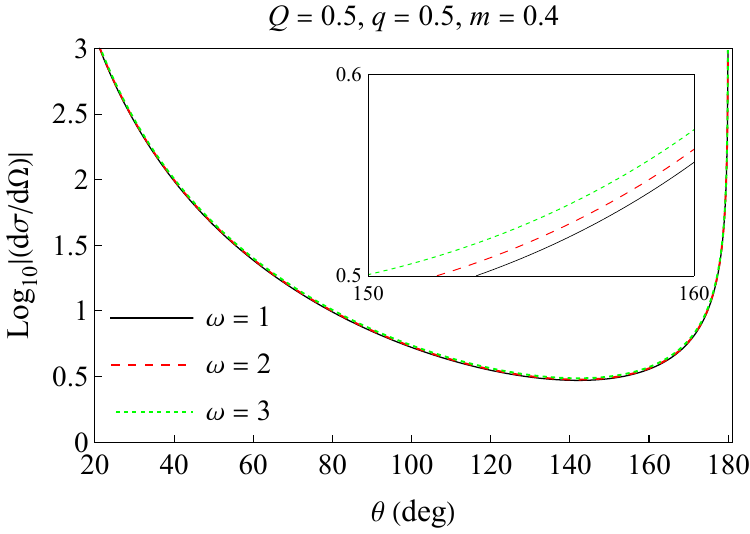}
		\end{tabular}
		\begin{tabular}{cc}
			\includegraphics[width=0.5 \textwidth]{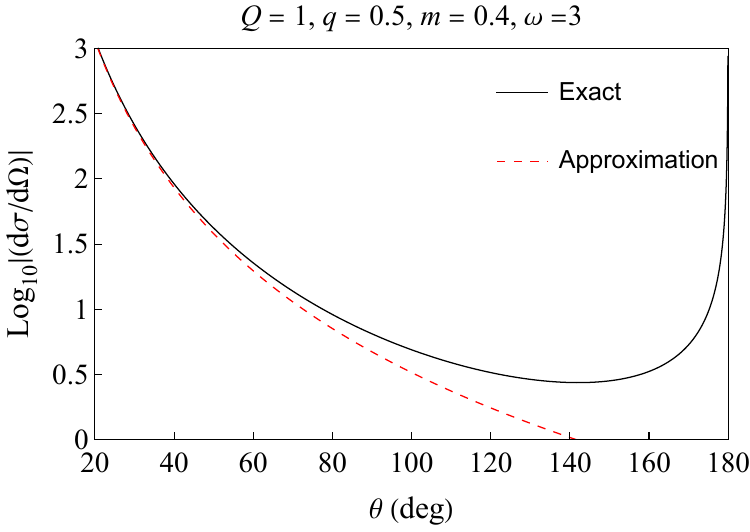}
		\includegraphics[width=0.5\textwidth, height=0.25\textheight]{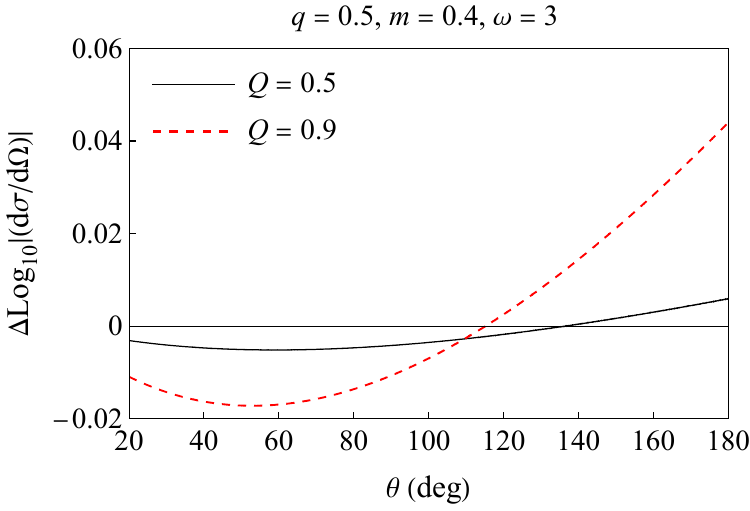}
		\end{tabular}
		\caption{	
			The classical scattering cross section of charged Horndeski BHs for different values of $Q$ (top-left), $q$ (top-right), $m$ (center-left), and $\omega$ (center-right).
			Comparison between the exact and approximate results \eqref{eq:wcsdiff} of classical scattering cross section in a charged Horndeski BH (bottom-left).  The bottom-right plot is the RN spacetime cross section subtracted by that of the charged Horndeski spacetime.}	\label{geo}	
	\end{figure}

In Fig. \ref{rc-bc}, the left and right axes of both plots show respectively the value of $b_c$ and $r_c$ obtained from the above equations in a charged Horndeski spacetime with $Q=0.5$ for a charged timelike signal with $q=0.5$ for varying $\omega$ and $m$. In comparison, the corresponding $b_c$ and $r_c$ for an RN BH with the same $Q$ is also given. It is seen from the left plot that as the particle energy increases and becomes relativistic ($m<\omega\lesssim 2m$), the critical $r_c$ and impact parameter $b_c$ decrease until minimal values. This is intuitively expected because a faster-moving particle requires a closer radial distance in order to be captured. However, as the energy continues to increase so that the particle becomes ultra-relativistic, we see that both $r_c$ and $b_c$ increase, for both the charged Horndeski and the RN BH spacetimes. We confirm that this is an electric effect because in the Schwarzschild spacetime case, both $b_c$ and $r_c$ decrease monotonically as $\omega$ increases. Note that this increase of $b_c$ and $r_c$ (in the RN spacetime) was not detected in Ref. \cite{Zhou:2022dze} and is important in explaining the behavior of the scattering cross section in Fig. \ref{geo}.
From the right plot of Fig. \ref{rc-bc}, we observe that both $r_c$ and $b_c$ in both the charged Horndeski and RN spacetimes are monotonically increasing as $m$ increases while fixing $\omega$. Similar to the decreasing part in the left plot, we also attribute this behavior to the gravitational origin. i.e., heavier particles with the same $\omega$ can be trapped even at a larger impact parameter.

When $q=0$, the equations determining $r_c$ and $b_c$ degenerate to
	\begin{align}
		2\left[1+\left(v^2-1\right)	A(r_c)\right]=r_c\frac{A'(r_c)}{A(r_c)},\label{rc1}\\
		b^2_c=\frac{r^2_c \left[1+\left(v^2-1\right)	A(r_c)\right]}{v^2  A(r_c)},\label{bc1}
	\end{align}
respectively.   
It is easy to obtain the results for $r_c$ and $b_c$ from the above pair of equations in the charged Horndeski spacetime, especially for null geodesics when $m=0$, i.e., $v=1$. 
The classical absorption cross section, known as the geometric cross section, is defined as \cite{Wald2010}
	\begin{align}\label{gcs}
		\sigma_{\text{gcs}}=\pi b_c^2.
	\end{align}

Defining $u=1/r$,  Eq. \eqref{radial1} can be rewritten as a differential equation of $u(\phi)$ as
	\begin{align}\label{orbiteq}
		\left(\frac{\dd u}{ \dd \phi}\right)^2&=\frac{\left(1+\frac{\sqrt{1-v^{2}}}{m} q A_{0}(u)\right)^{2}}{b^2v^2A(u) B(u)}-\frac{1}{B(u)}\left(\frac{1-v^2}{b^2v^2}+u^2\right)\\ \nonumber
		&\equiv  h(u),
	\end{align}
where $h(u)$ is defined as a function of the form of the right-hand side. 
The deflection angle of the signal is then \cite{Collins:1973xf}
	\begin{align}\label{Theta_b1}
		\Delta\theta(b)=2\int_{0}^{u_0}\frac{\dd u}{\sqrt{h(u)}}-\pi,
	\end{align}
where $u_0=1/r_0$ satisfies the condition $h(u_0)=0$.
Using the perturbative approach,  Xu et al. have calculated in the weak deflection limit the deflection angle of charged particles in the charged Horndeski spacetime as \cite{Xu:2021rld}
		\begin{align}\label{deflection}
			\Delta\theta(b)  &\approx\left(1+\frac{1}{v^2}-\frac{\tilde{q}\tilde{Q}\sqrt{1-v^2}}{v^2}\right)\frac{2M}{b}\\	\nonumber
			&+\frac{\pi}{4} \left[3\left(1+\frac{4}{v^2} \right)-\left(1+\frac{1}{v^2}\right)\frac{\hat{Q}^2}{2}-\frac{12 \hat{q} \hat{Q} \sqrt{1-v^2}}{v^2}+2\left(\frac{1}{v^2}-1\right)\hat{q}^2 \hat{Q}^2 \right]\frac{M^2}{b^2},  
		\end{align}
where $\tilde{q}\equiv\frac{q}{m}$ and $\tilde{Q}\equiv \frac{Q}{M}$. When $\tilde{q}=0$ and $v=1$, we obtain the weak deflection angle of light in this spacetime \cite{Wang:2019cuf}
	\begin{align}\label{weakdeflection}
		\Delta\theta(b) \approx\frac{4 M}{b}+\frac{15 M^2 \pi -\pi Q^2}{4 b^2}.
	\end{align}
The classical differential scattering cross section is related to the relation $\theta(b)$ by \cite{Collins:1973xf}
	\begin{align}\label{cdscs}
		\frac{\dd \sigma_\mathrm{C}}{ \dd \Omega}= \sum_{N} \frac{b}{\sin\theta}\bigg|\frac{ \dd b}{ \dd \theta}\bigg|,
	\end{align}
where $\theta$ is the scattering angle that can be linked to $\Delta\theta$ by the simple relation $\theta = |\Delta\theta - 2N\pi|$ with $N = 0, 1, 2, \dots$ representing the number of loops the particles go around the BH before escaping to infinity. For later usage, we will invert the function $\Delta\theta(b)$ in Eq. \eqref{Theta_b1} to a function $b(\Delta\theta)$ and then further convert it to a function of the scattering angle $\theta$,  denoting this new function as $b(\theta)$. In particular, 
for small deflection/scattering angles, by using Eqs. \eqref{deflection} and \eqref{cdscs}, one can obtain the analytical expression for $\displaystyle\frac{\dd \sigma_\mathrm{C}}{\dd \Omega}$
at small angles 
\begin{align} \label{eq:wcsdiff}
	\frac{\dd \sigma_\mathrm{C}}{\dd \Omega}\simeq &\frac{4M^2}{\theta^4}\left(1 + \frac {1} {v^2} - \frac{ \sqrt{1-v^2} \tilde {q} \tilde{Q}} {v^2}\right)^2\\  \nonumber
	&-\frac{\pi M^2} {8 v^2 \theta^3}\left\{-6\left(4+v^2\right)+24 \sqrt{1-v^2}\tilde{q}\tilde{Q} +\left[1 + v^2+4\left(1 - v^2\right)\tilde {q}^2 \right] \tilde {Q}^2 \right\}.
\end{align}
Note that in the semiclassical limit, the energy $E$ is related to the wave frequency $\omega$ by $E=\hbar\omega$ and the angular momentum $L$ to the angular quantum number $l$ by $|L|=\hbar\left|l+\frac{1}{2}\right|$. Therefore,  the $v$ here can also be converted to $\omega$ by $v=\sqrt{1-m^2/\omega^2}$.

To investigate the impact of the parameters $Q$, $q$, $m$, and $\omega$ on $\displaystyle\frac{\dd \sigma_\mathrm{C}}{\dd \Omega}$, we plot the latter against the scattering angle for different parameter values in Fig. \ref{geo}. Furthermore, it is worth emphasizing that we use $M=1$ throughout the entire calculation so that all quantities with mass dimensions are effectively scaled by $M$, and the field mass $m$ and charge $q$ are scaled by $M^{-1}$. Since our study treats the scattering field as a perturbation, it is expected that the field mass $m$ and charge $q$ are small so that the existence and propagation of the field in the spacetime do not change the background charged Horndeski spacetime. The field parameter values we chose in the plotting can always be made to satisfy this condition by properly adjusting $M$.   The top panel of Fig. \ref{geo} display $\displaystyle\frac{\dd \sigma_\mathrm{C}}{\dd \Omega}$ in two distinct sets: (i) at $q=0$, $m=0.4$, $\omega=3$, with varying values of $Q$ (top-left); (ii) at $Q=0.5$, $m=0.4$,  $\omega=3$, with varying values of $q$ (top-right).   This shows how $\displaystyle\frac{\dd \sigma_\mathrm{C}}{\dd \Omega}$ is affected by the charge of the BH and field in two scenarios: one with and one without electromagnetic interaction.  We see that the larger the value of $Q$, the smaller the $\displaystyle\frac{\dd \sigma_\mathrm{C}}{\dd \Omega}$. This is consistent with the known fact that larger $|Q|$ will shrink the BH shadow in charged spacetimes \cite{Pang:2018jpm}.  From the top-right panel of Fig. \ref{geo}, we observe that  $\displaystyle\frac{\dd \sigma_\mathrm{C}}{\dd \Omega}$ decreases with the increase of the field charge.  This is similar to the behavior of the critical impact parameter $b_c$ under the influence of $q$ in RN spacetime \cite{Zhou:2022dze}. Additionally, we demonstrate the impact of the field mass $m$ on $\displaystyle\frac{\dd \sigma_\mathrm{C}}{\dd \Omega}$ for fixed $Q=q=0.5$ and $\omega=3$ (center-left), and  the effect of incident frequency $\omega$ for fixed $Q=q=0.5$ and $m=0.4$ (center-right). We find that both the increase of the field mass and frequency lead to a slight increase in $\displaystyle\frac{\dd \sigma_\mathrm{C}}{\dd \Omega}$.  Both these two features are understandable from the effect of $m$ and $\omega$ on the critical impact parameter $b_c$ as shown in Fig. \ref{rc-bc}. Intuitively, the increase of the cross section as $m$ increases is because a larger $m$ with fixed $\omega$ (equivalently $E$) will cause a smaller velocity of the incident signal and therefore increase the scattering cross section. However, the increase in cross section as $\omega$ increases in the given range in the plot is due to the electric effect, since this does not happen in the scattering of the same wave by a Schwarzschild BH.   Additionally, in the bottom-left, we compare the exact and approximate result \eqref{eq:wcsdiff} at small angles. It is observed that Eq. \eqref{eq:wcsdiff} approximates the exact values very well. Finally, in the bottom-right panel, we compare the effect of BH charge of the charged Horndeski and RN BHs on $\displaystyle\frac{\dd \sigma_\mathrm{C}}{\dd \Omega}$. In this plot,  we present the difference between the RN and charged Horndeski BHs scattering cross sections for the same BH charge parameters. We can see that the classical scattering cross sections of RN BH are always lower than those of charged Horndeski BH at small scattering angles.

\subsubsection{Glory scattering} \label{analysis_glory}

Because scattered waves with different angular momenta exhibit interference phenomena similar to those in optics when the scattering angles approach $\pi$, the combination of the scattered wave and the incident wave produces a bright spot or halo known as a glory. An analytical formula of the glory scattering cross section $\displaystyle\frac{\dd\sigma_\mathrm{G}}{\dd\Omega}$ was provided in Ref. \cite{Ford:2000uye} in the semiclassical approximation to describe these characteristics. Based on them, for the massless wave of frequency $\omega$, Zhang and DeWitt-Morette  \cite{Zhang:1984vt} derived $\displaystyle\frac{\dd\sigma_\mathrm{G}}{\dd\Omega}$  that can be directly extended to massive waves, which reads
	\begin{align}\label{glory}
		\frac{\dd \sigma_\mathrm{G}}{\dd \Omega}\simeq 2\pi\omega v b^2_g\Big|\frac{\dd b}{\dd \theta}\Big|_{\theta\simeq \pi}[J_{2s}\left(\omega  v b_g\sin\theta\right)]^2,
	\end{align}
where $b_g=b(\pi)$ represents the glory impact parameter and $J_{2s}$ is the Bessel function of the first kind, with $s$ being the helicity of the considered wave. Here, $s=0$ because we are dealing with a scalar wave. Previously, we mentioned that there are multiple deflection angles $\Delta\theta$'s that correspond to the same scattering angle $\theta$. Therefore, $b_g$ also has multiple corresponding values. All the backward scattering of massive waves close to $\pi$ contributes to glory scattering, but the contribution can be neglected when the scattered waves have undergone more than 1 loop before going to infinity. In other words, in the numerical calculation of glory scattering, we only consider the contribution of the case where $N=0$.
In Fig. \ref{sca1}, we will use the approximate  \eqref{glory} to obtain the glory scattering cross section and compare it with that obtained using the numerical method.

\section{Absorption and scattering cross section results and analysis}\label{results}

In this section, we discuss the absorption and scattering cross sections obtained by solving the Schr$\ddot{\textrm{o}}$dinger-like Eq.  \eqref{Scheq} with the boundary conditions  \eqref{solution}.  We then compare the numerical results of the cross sections with those of the approximation.

In order to obtain the reflection coefficient, we numerically integrate this second-order differential equation \eqref{Scheq} from the near-horizon region to far away using the fourth-order Runge-Kutta method, and then fit the obtained results with the boundary condition at infinity (see Ref. \cite{Dolan:2009zza} for more detail). Because the sum of Eq.  \eqref{sa} has poor convergence for $\theta\approx0$, we also used the so-called reduced series (see Refs. \cite{Yennie:1954zz,Dolan:2006vj} for more details) to deal with this shortcoming.

\subsection{Absorption cross section}\label{Absorption}

\begin{figure}[htp]
	\centering
	\begin{tabular}{cc}
		\includegraphics[width=0.5 \textwidth]{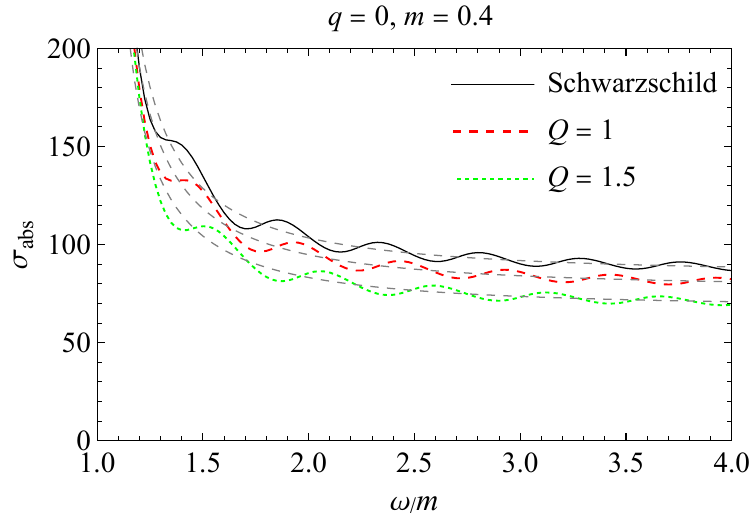}
		\includegraphics[width=0.5 \textwidth]{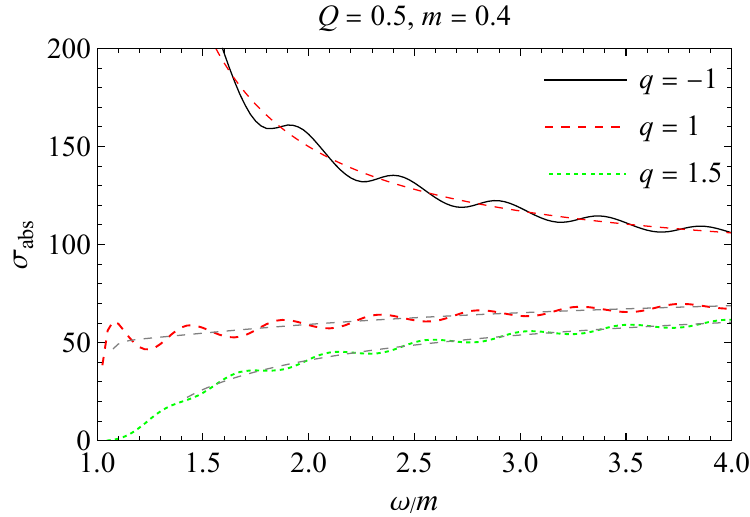}
	\end{tabular}
	\caption{ The total absorption cross section of charged Horndeski BHs for different values of $Q$ (left) and $q$ (right). The gray dashed lines represent the classical result in the high-frequency limit.}
		\label{abs1}
\end{figure}

Fig. \ref{abs1} shows the $\sigma_{\mathrm{abs}}$ for charged Horndeski BH with an uncharged massive scalar field (left) and a charged massive scalar field  (right), where the field mass is the same ($m = 0.4$). The values of the BH charge are set to $Q = 0,\, 1,\,1.5$ in the left plot, where the case $Q = 0$ corresponds to the Schwarzschild BH case. It can be seen that for an uncharged scalar field, the $\sigma_{\mathrm{abs}}$ decreases with higher values of BH charge and always tends to infinity when $\omega/m \rightarrow 1$. Similarly, when electromagnetic fields exist, the $\sigma_{\mathrm{abs}}$ also decreases as the field charge increases, but in the limit $\omega/m \rightarrow 1$, it approaches a smaller value when $qQ>0$. These properties are anticipated due to the fact that Lorentz repulsion counteracts the attractive nature of gravitational force.  Furthermore, the numerical results (colored lines) exhibit oscillatory behavior around the classical ones, $\pi b_{c}^{2}$  (gray lines),  in the high-frequency limit.  This indicates that we no longer need to visualize $\sigma_{\mathrm{abs}}$ vs $m$ anymore.  We can also obtain the fact from the right panel in Fig. \ref{rc-bc} that, while keeping other parameters fixed, an increase in field mass leads to a larger $\sigma_{\mathrm{abs}}$.   In other words, a heavier field tends to be absorbed more easily than a lighter one with the same velocity (or energy/mass ratio).

\begin{figure}[htp]
	\centering
	\includegraphics[width=0.5 \textwidth]{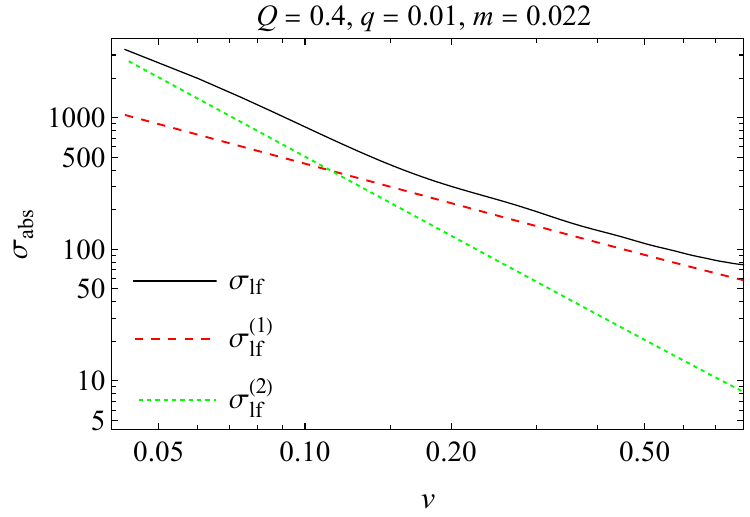}
	\caption{ Comparison of the numerical partial absorption cross section with the approximate analytical results, i.e., Eqs. \eqref{low_abs1} and \eqref{low_abs2} for $l=0$ with $v_{c}\approx0.11$}. \label{low-abs}

\end{figure}
	
In Fig. \ref{low-abs}, we compare the numerical partial absorption cross section with	the approximate analytical results in the low-frequency limit, namely, Eqs. \eqref{low_abs1} and \eqref{low_abs2}. We can see that there is a switch in the numerical result from $\sigma_{lf}^{(1)}$ to $\sigma_{lf}^{(2)}$ when $v_{c}\approx0.11$.
	
	\begin{figure}[htp]
		\centering
		\begin{tabular}{cc}
   	    \includegraphics[width=0.5 \textwidth]{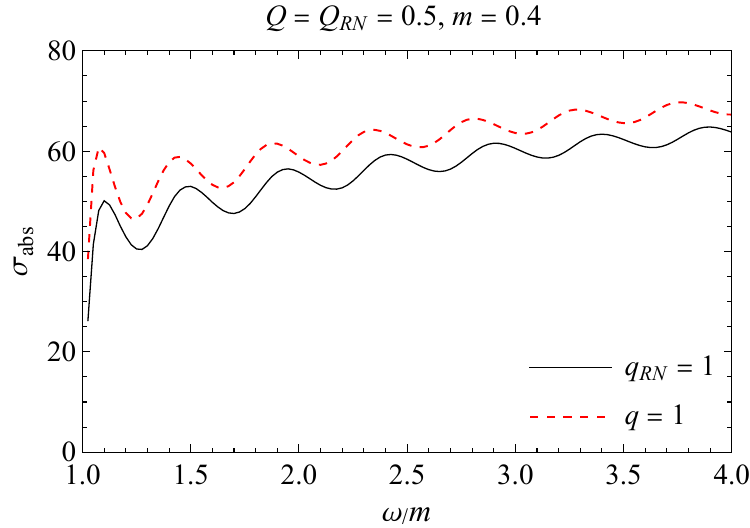} 
		\end{tabular}
		\caption{ 
				Comparison between the absorption cross section of a charged massive scalar field in charged Horndeski and RN BHs.}	\label{com_abs_rn}
	\end{figure}
	
Fig. \ref{com_abs_rn} compares the absorption cross sections of the charged Horndeski and RN BHs with fixed parameters $Q=Q_{RN}=0.5$, $m=0.4$ and $q=1$.  Besides, the right plot of Fig. \ref{rc-bc} has shown that $b_c$ ($\pi b_{c}^{2}$) of a charged Horndeski BH is larger than that of an RN BH for the same $m$. Moreover, as $m$ increases, $b_c$ of both spacetimes increases while their relative size relation still holds.
Therefore, we do not have to plot the cross-section to compare the effect of the mass $m$ on the absorption cross section in these two BH spacetimes.  From this analysis and Fig. \ref{com_abs_rn}, it is evident that scalar waves propagating in the charged Horndeski BH  are more absorbed than those in the RN BH for the same values of parameters.    This conclusion can be seen in the definition and inequality \eqref{condition} of the BH charge $Q$ in charged Horndeski and RN  BHs. In other words,   the range of values of BH charge $Q$ in charged Horndeski is larger than that of  BH charge $Q$ in RN BHs. Similar phenomena happen in other charged spacetime too.   We have calculated the impact parameter $b_c$ for various BHs, including  Ay\'on-Beato \& Garc\'ia (ABC) BHs, Bardeen BHs, RN BHs and Hayward BHs with their respective extremal charges: $Q_{ABC}\approx0.63M$,  $Q_{Bardeen} \approx 0.77M$,  $Q_{RN} = 1M$ and  $Q_{Hayward} \approx 1.06M$  (seeing the Table 1 of Ref. \cite{Wan:2022vcp}).  It is found that the relation $b_c^{Hayward} > b_c^{RN}> b_c^{Bardeen}> b_c^{ABC}$ agrees well with the relative ordering of the extreme $Q$ values in the corresponding spacetimes. Consequently, the absorption cross sections in these spacetimes under the same $Q$ will also be ordered as above.

\subsection{Differential scattering cross section}\label{scattering}

\begin{figure}[H]
		\centering
		\begin{tabular}{cc}
			\includegraphics[width=0.5 \textwidth]{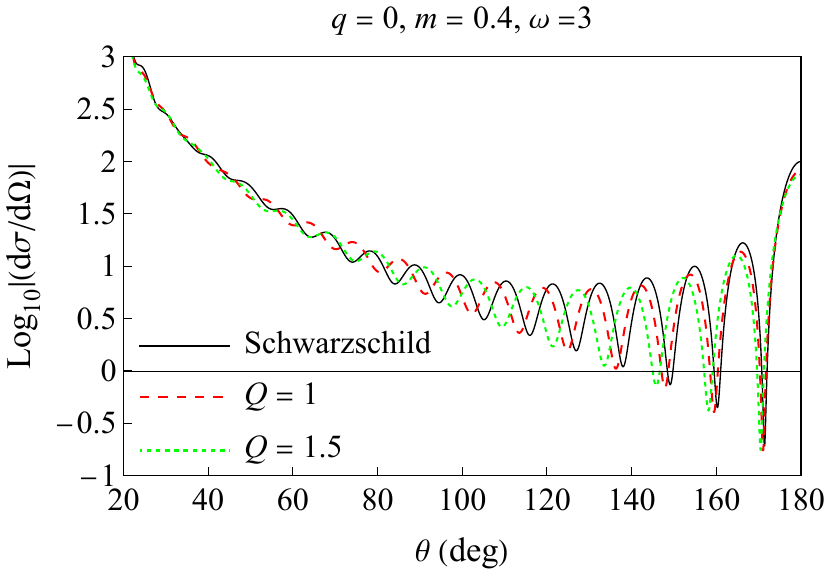}
			\includegraphics[width=0.5 \textwidth]{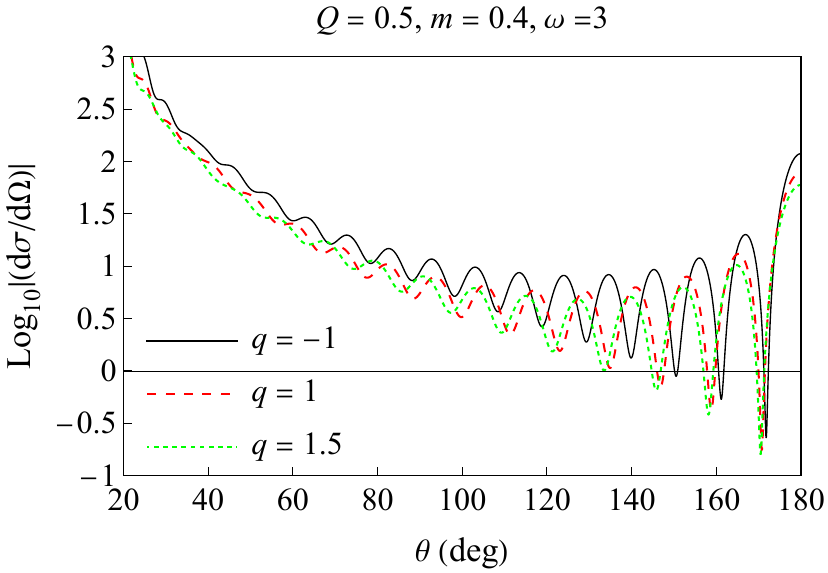}
		\end{tabular}
			\begin{tabular}{cc}
			\includegraphics[width=0.5 \textwidth]{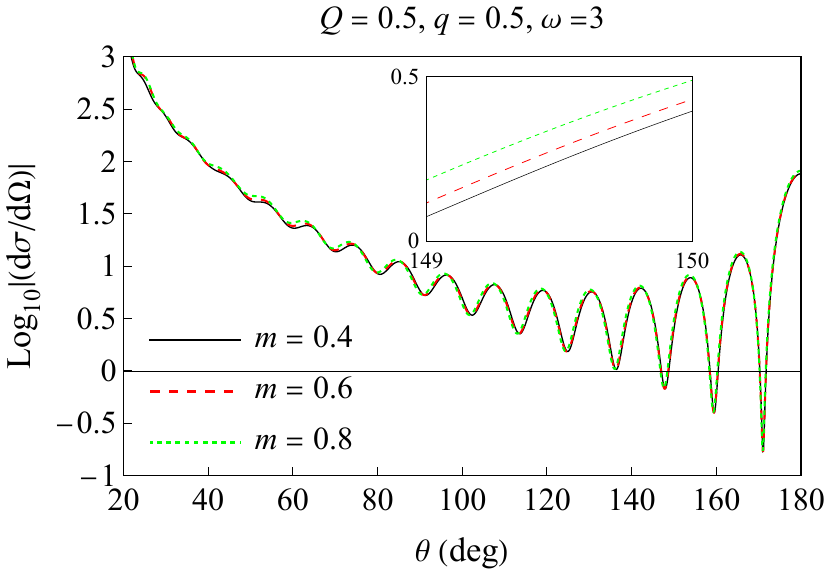}
			\includegraphics[width=0.5 \textwidth]{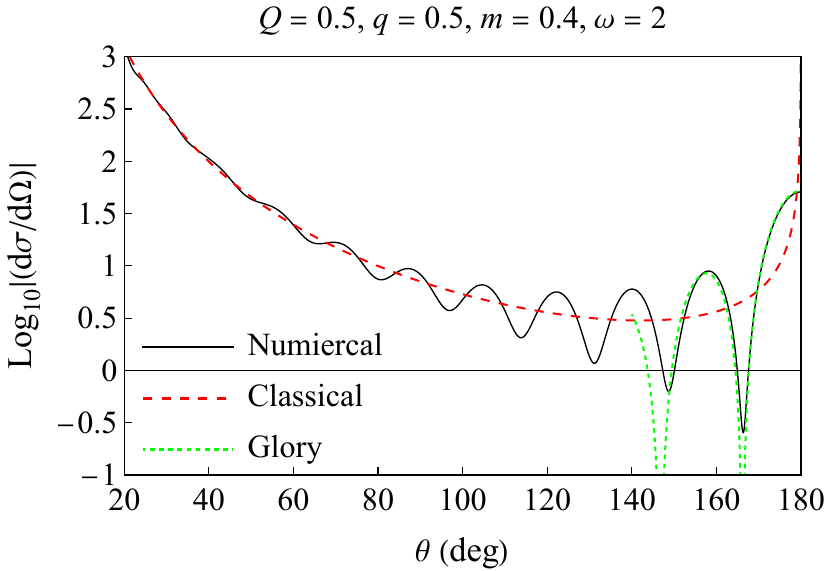}
		\end{tabular}
		\caption{ The differential scattering cross section of charged Horndeski BHs for different values of $Q$ (top-left), $q$ (top-right) and  $m$ (bottom-left). Comparison of the differential scattering cross section of charged Horndeski BHs obtained by geodesic scattering, the glory scattering, and numerical results (bottom-right).	}\label{sca1}
	\end{figure}

Similar to the absorption cross section, in Fig. \ref{sca1}, we consider the differential scattering cross section using the same parameter settings. In the top panel, we study the differential scattering cross sections of an uncharged and a charged massive scalar wave in two different parameter settings: (i) with $q=0$, $m=0.4$, $\omega=3$, showing the effect of varying $Q$ (top-left), and (ii) with $Q=0.5$, $m=0.4$, $\omega=3$, showing the effect of varying $q$ (top-right).   We find that the stripe width of the differential scattering cross section becomes larger as the BH charge value increases for an uncharged massive scalar field.  Furthermore,  when exploring the effect of varying $q$, it is evident that an increase in the positive field charge leads to a decrease in the scattering intensity and an expansion in the width of the interference fringes. This indicates that the rise in the field charge will significantly decrease the scattered flux and exhibit an opposite effect compared to the gravitational interaction when $qQ > 0$.

The center-left plot in Fig. \ref{sca1} displays the effects of field mass on the differential scattering cross sections, with the parameters set to  $Q=q=0.5$ and $\omega=3$. We note that an increase in field mass results in a slight widening of the fringe width and an increase in the magnitude of the oscillations. This observation is consistent with conclusions drawn from the scattering of Dirac fermions by Schwarzschild BHs \cite{Cotaescu:2014jca}.  In addition, on the bottom-right plot of Fig. \ref{sca1}, we present a comparison between geodesic scattering Eq. \eqref{cdscs}, the glory scattering Eq. \eqref{glory}, and the numerical results obtained from the partial wave method Eq. \eqref{dscs} for fixed parameters $Q=q=0.5$, $m=0.4$ and $\omega=2$.  We observe that the glory scattering result describes the interference fringes of scalar waves with scattering angles close to $\pi$ with high precision, especially for $\theta>160^{\circ}$. On the other hand, geodesic scattering resembles the average value of the partial wave results which oscillate around the former one.  Finally, by comparing with the case $Q=q=0.5$, $m=0.4$ and $\omega=3$ on the bottom-left plot, we find that the interference fringe widths become narrower, and the oscillation amplitudes of the scattering cross section increase with higher wave frequency. 
 
	\begin{figure}[htp]
	\centering
	\begin{tabular}{cc}
		\includegraphics[width=0.5 \textwidth]{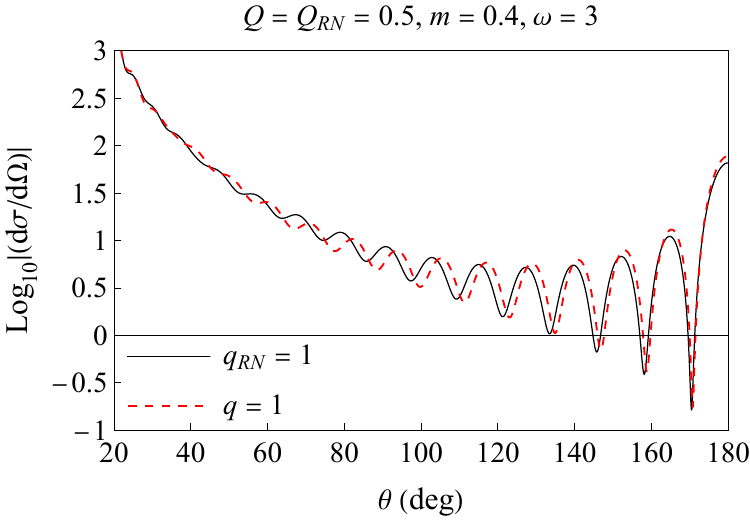}
		\includegraphics[width=0.5 \textwidth]{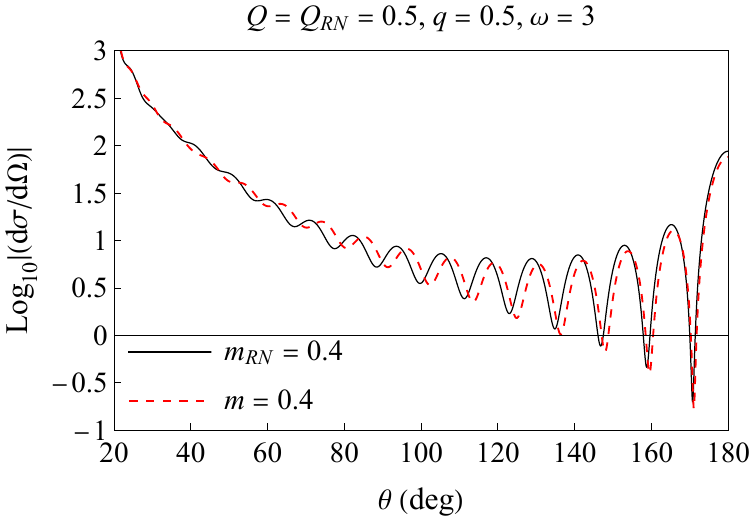}
	\end{tabular}
	\caption{ Comparison between the scattering cross sections of a charged massive scalar field in the charged Horndeski and RN  BHs for the same  $q$ (right) and $m$ (left).}	\label{com_sca_rn}
\end{figure}
 
 A comparison of the differential scattering cross sections in charged Horndeski and RN BHs is depicted in Fig. \ref{com_sca_rn} for the same values of $q$ and $m$.  We find that the width of the interference fringes in RN BHs is larger than that in charged Horndeski BHs for the same value of $q$ and $m$.

\section{Superradiance}\label{superradiance}

A charged BH can also extract charge and mass from the BH, a phenomenon known as the superradiance effect. As seen from Eq. \eqref{rtf}, superradiant scattering also occurs for the charged Horndeski BH when $k_h<0$, i.e., when
\begin{align}
m<\omega<-q A_0(r_h). \label{eq:condition}
\end{align}
In this section, we aim to investigate the superradiance in the absorption and scattering cross sections.

	\begin{figure}[htp]
		\centering
		\begin{tabular}{cc}
			\includegraphics[width=0.5 \textwidth]{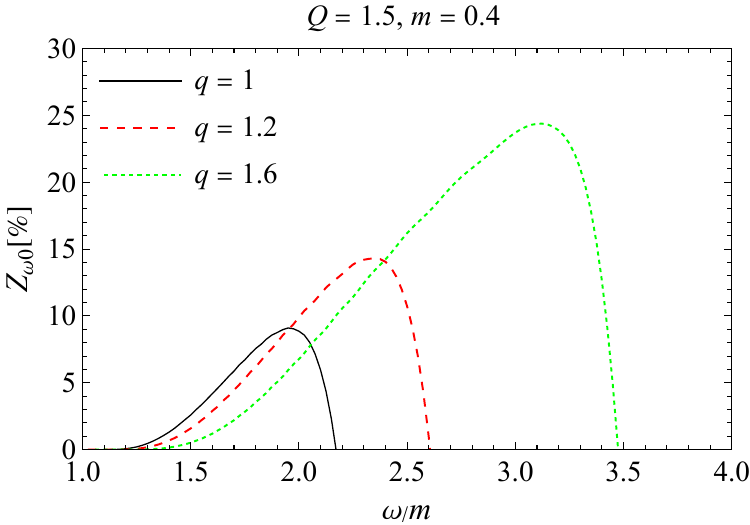}
			\includegraphics[width=0.5 \textwidth]{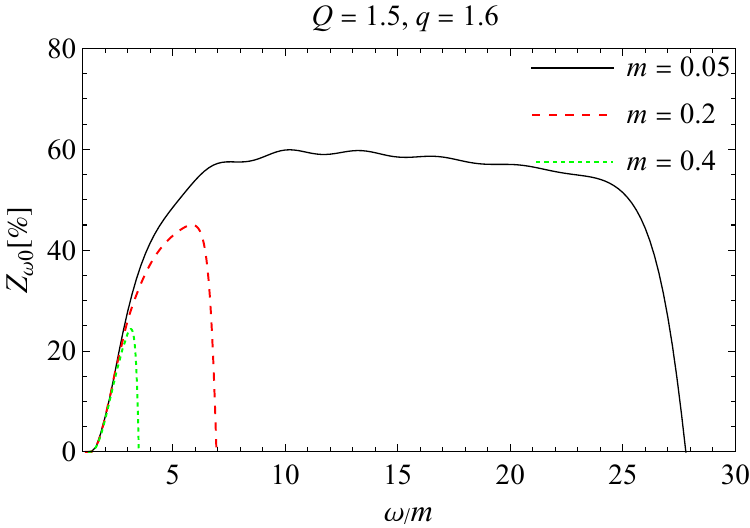}
		\end{tabular}
		\caption{ The amplification factor of charged Horndeski BHs for the monopole wave $l=0$  for different values of   $q$ (left) and $m$ (right).}	\label{super_am}
	\end{figure}
 
To quantify superradiance more effectively, one can define an amplification factor $Z_{\omega l}$ as
	\begin{align}\label{am}
		Z_{\omega l}=|R_{\omega l}|^2 -1.
	\end{align}
When $Z_{\omega l}>0$, superradiance occurs around the BH and otherwise, there is no such phenomenon. In Fig. \ref{super_am},  we present the amplification factor as a function of $\omega/m$ for the monopole wave $l=0$ and fixed parameter $Q=1.5$. The effects of the field charge and field mass are shown respectively in the left and right plots. It is evident from the left plot that as the Lorentz repulsion force around the BH increases, both the amplification factor and the corresponding frequency range in which the superradiance occurs, i.e. $\omega/m\in(1,-qA_0(r_h)/m)$, increase. In other words, a more repulsive incoming wave can extract more energy from the BH. On the other hand, if the condition $m<\omega<-q A_{0}(r_h)$ is met, then the superradiance can also be affected by the field mass for a massive scalar field. Thus, we observe in the right plot that as the field mass decreases, the amplification factor increases in general while the range for the superradiance $\omega/m\in(1,-qA_0(r_h)/m)$ widens. This implies that a scalar wave with a smaller mass makes it easier to extract energy from the charged Horndeski BH. 

\begin{figure}[htp]
		\centering
		\begin{tabular}{cc}
		\includegraphics[width=0.5 \textwidth]{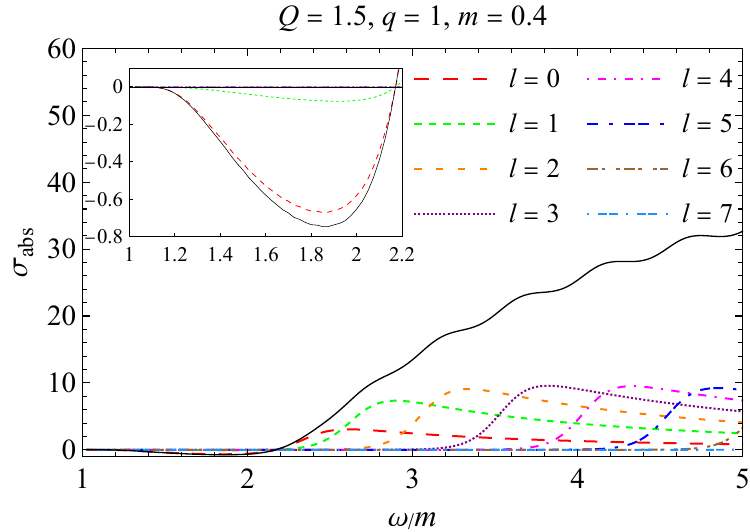}
		\end{tabular}
		\caption{The partial (colorful lines) and total (black line)  absorption cross sections of charged Horndeski BH with parameters  $Q=1.5$, $q=1$ and $m=0.4$.
		}	\label{super_abs_sca}
\end{figure}

The above analysis concerns the amplification factor (reflection coefficient). We now extend the discussion of the superradiance phenomenon to the absorption cross section with the presence of repulsive electromagnetic interaction. Fig. \ref{super_abs_sca} shows the absorption cross section as a function of $\omega/m$ for $Q=1.5$, $q=1$, and $m=0.4$. A peculiarity, compared to the right plot in Fig. \ref{abs1}, is that the total and partial absorption cross sections for the angular modes $l=0$ and $l=1$ exhibit negative values, which is consistent with the absorption of charged scalar wave by an RN BH \cite{Benone:2015bst}. Comparing to the left plot of Fig. \ref{super_am}, we see that the frequency range for the negative absorption cross section, i.e., roughly $\omega<2.2m$, is exactly where the superradiance happens in the amplification factor for $q=1$. This number, in turn, equals exactly 
\begin{align}
    \omega_c=-q A_0(r_h)=4\left(\frac{8}{3}-\sqrt{6}\right)\approx 2.2,
\end{align}
where for the current parameter settings, $\displaystyle r_h=\frac{3}{8}\left(2+\sqrt{6}\right)$ is found by solving Eq. \eqref{A,B} and Eq. \eqref{eq:A0} is used for $A_0(r_h)$.
	
When superradiance occurs,  charged scalar waves are amplified by the BH and some scalar waves that would have been absorbed by the BH may escape to infinity \cite{Glampedakis:2001cx}. In other words, for superradiance, the scattering flux may also be enhanced, and the geodesic analysis in Subsec. \ref{analysis-high} for the differential scattering cross section becomes inapplicable due to the complex impact parameter. However, when using the partial wave approach to calculate the scattering cross section, no similar limiting condition exists as long as the superradiance condition Eq. \eqref{eq:condition} is satisfied. 

\begin{figure}[htp]
		\centering
		\begin{tabular}{cc}
			\includegraphics[width=0.5 \textwidth]{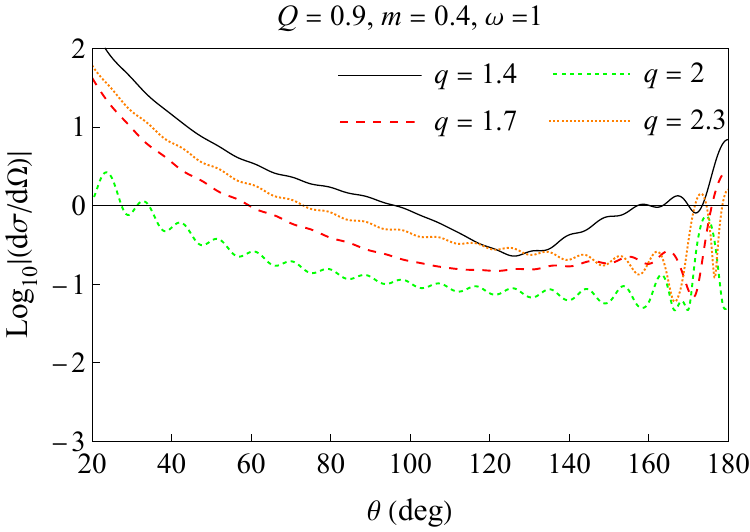}
			\includegraphics[width=0.5 \textwidth]{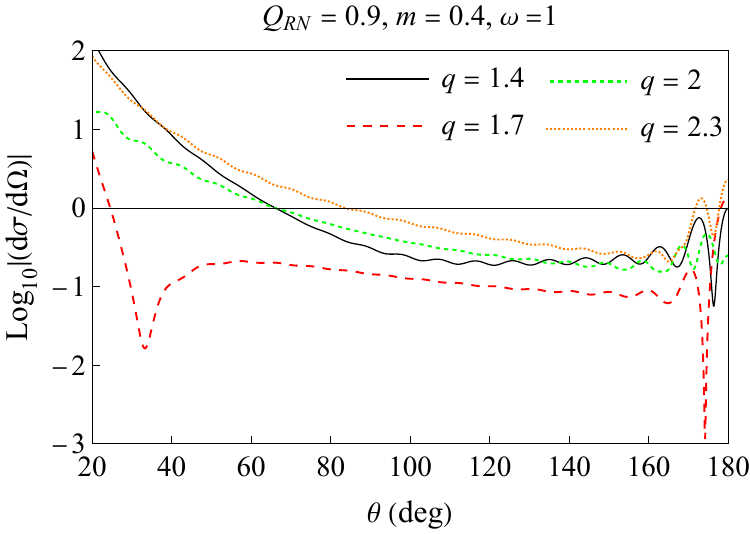}
		\end{tabular}
		\caption{The differential scattering cross sections of charged Horndeski BH and RN BH with fixing the parameters  $Q=Q_{\rm{RN}}=0.9$, $m=0.4$ and $\omega=1$ varying the field charge.
		}	\label{super_sca}
	\end{figure}

To investigate the change in the differential scattering cross section associated with superradiance, Fig. \ref{super_sca} displays the scattering cross section as a function of $\theta$ for various field charges $q$, while maintaining the parameters $Q=0.9$, $m=0.4$, and $\omega=1$ constant. 
For this parameter setting,  the superradiance condition $\omega=1<-qA(r_h)$ implies that when the field charge is larger than $q_{c}\approx2.1$ for the charged Horndeski BH and $q_{c}^{RN} \approx 1.6$ for the RN BH, superradiance will happen.
 In the top-right plot of Fig. \ref{sca1} where the superradiance condition is not met, we have observed that the greater the Lorentz repulsion force, the wider the interference fringes and the smaller the scattering intensity. 
Here we see that as $q$ gradually increases, the scattering cross section continues to decrease, although the oscillations become more irregular, especially near the backward direction. When $q$  becomes greater than $q_c$ (the $q=2.3$ line in the left plot and the $q=1.7,\,2,\,2.3$ lines in the right plot), the scattering intensity increases as $q$ increases. This is another indication that superradiance enhances the scattering flux. It is also worth noting that for the superradiance case, the maxima of the oscillation patterns do not necessarily occur at $\theta=180^\circ$ anymore as seen from the $q=2$ line of the RN BH case (right plot).

\section{Conclusion and discussion}	\label{conclusion}

 We computed the absorption and scattering cross sections of a charged massive scalar wave impinging on the charged Horndeski BH using the approximation or classical geodesic approach as well as the partial wave method. We compared the numerical results with approximate analytical results, finding them to be in good agreement.

To understand how the BH's charges influence the absorption and scattering cross sections, we compared our results for a charged Horndeski BH with those of a Schwarzschild BH. Our findings indicate that charged Horndeski BHs with smaller charges can absorb more neutral scalar waves, and the fringe width of their scattering cross sections is narrower. This is consistent with the effect of the BH charge on the photon sphere size. 
 
We then investigated how the Lorentz force impacts the absorption and scattering cross sections in the charged Horndeski BH.  When $q Q>0$, the repulsive force results in a decrease in absorption cross sections and an increase in the fringe widths of differential scattering cross sections as the values of $q$ increase. This is in accordance with the classical picture that a Lorentz repulsion (or attraction) force can result in a smaller (or larger) photon and particle sphere (i.e. critical radius $r_c$). If one of the signs of the BH charge and the field charge is opposite, then the effects described above will be reversed. Moreover, the comparison of charged Horndeski and RN BHs shows that the absorption cross section in the charged Horndeski spacetime is larger than in the RN case for the same charge parameters.  For the scattering cross section, it is found that the fringe widths in the RN BH are wider than those in the charged Horndeski BH for the same parameters.

For superradiance, we analyzed the effect of the field charge and field mass on the amplification factor and found that an increase in the field charge or a decrease in field mass leads to an increase in the amplification factor. We then calculated the absorption and differential scattering cross sections when the superradiance happens, i.e., when $m<\omega<-q A_{0}(r_h)$. We see that both the total and partial absorption cross sections become negative due to the Lorentz repulsion force. And the differential scattering cross section is enhanced by superradiance when $m<\omega<-q A_{0}(r_h)$. All of these imply that charged planar waves can be amplified by the charged Horndeski BH. These conclusions highlight the significant role of the repulsive Lorentz force in superradiance in the scattering problem for a charged scalar wave.

For future directions, we note that when an uncharged scalar wave interacts with the Kerr BH, the superradiance resulting from the BH's rotation has a negligible effect. Our next focus will be the impact of field charge, as well as the BH rotation, on the scattering problem with charged massive scalar waves interacting with a charged and rotating BH.

	\noindent

\begin{thebibliography}{100}

	\bibitem{Horndeski:1974wa} G.~W.~Horndeski,
	Int. J. Theor. Phys. \textbf{10} (1974), 363-384
	doi:10.1007/BF01807638.
	
	%

	\bibitem{Amendola:1993uh} L.~Amendola,
	Phys. Lett. B \textbf{301} (1993), 175-182
	doi:10.1016/0370-2693(93)90685-B
	[arXiv:gr-qc/9302010 [gr-qc]].

	\bibitem{Saridakis:2010mf} E.~N.~Saridakis and S.~V.~Sushkov,
	Phys. Rev. D \textbf{81} (2010), 083510
	doi:10.1103/PhysRevD.81.083510
	[arXiv:1002.3478 [gr-qc]].
	
	
	
	

	\bibitem{Charmousis:2011bf} C.~Charmousis, E.~J.~Copeland, A.~Padilla and P.~M.~Saffin,
	Phys. Rev. Lett. \textbf{108} (2012), 051101
	doi:10.1103/PhysRevLett.108.051101
	[arXiv:1106.2000 [hep-th]].
	

	\bibitem{Maselli:2016gxk} A.~Maselli, H.~O.~Silva, M.~Minamitsuji and E.~Berti,
	Phys. Rev. D \textbf{93} (2016) no.12, 124056
	doi:10.1103/PhysRevD.93.124056
	[arXiv:1603.04876 [gr-qc]].

	\bibitem{Kobayashi:2019hrl} T.~Kobayashi,
	Rept. Prog. Phys. \textbf{82} (2019) no.8, 086901
	doi:10.1088/1361-6633/ab2429
	[arXiv:1901.07183 [gr-qc]].
	
	

	\bibitem{Galeev:2021xit} R.~Galeev, R.~Muharlyamov, A.~A.~Starobinsky, S.~V.~Sushkov and M.~S.~Volkov,
	Phys. Rev. D \textbf{103} (2021) no.10, 104015
	doi:10.1103/PhysRevD.103.104015
	[arXiv:2102.10981 [gr-qc]].
	
	
	

	\bibitem{Rinaldi:2012vy} M.~Rinaldi,
	Phys. Rev. D \textbf{86} (2012), 084048
	doi:10.1103/PhysRevD.86.084048
	[arXiv:1208.0103 [gr-qc]].
	
	

	\bibitem{Anabalon:2013oea} A.~Anabalon, A.~Cisterna and J.~Oliva,
	Phys. Rev. D \textbf{89} (2014), 084050
	doi:10.1103/PhysRevD.89.084050
	[arXiv:1312.3597 [gr-qc]].
	

	\bibitem{Cisterna:2014nua} A.~Cisterna and C.~Erices,
	Phys. Rev. D \textbf{89} (2014), 084038
	doi:10.1103/PhysRevD.89.084038
	[arXiv:1401.4479 [gr-qc]].
	

	\bibitem{Maselli:2015yva} A.~Maselli, H.~O.~Silva, M.~Minamitsuji and E.~Berti,
	Phys. Rev. D \textbf{92} (2015) no.10, 104049
	doi:10.1103/PhysRevD.92.104049
	[arXiv:1508.03044 [gr-qc]].
	
	

	\bibitem{Babichev:2016fbg} E.~Babichev, C.~Charmousis, A.~Leh\'ebel and T.~Moskalets,
	JCAP \textbf{09} (2016), 011
	doi:10.1088/1475-7516/2016/09/011
	[arXiv:1605.07438 [gr-qc]].
	

	\bibitem{Antoniou:2017hxj} G.~Antoniou, A.~Bakopoulos and P.~Kanti,
	Phys. Rev. D \textbf{97} (2018) no.8, 084037
	doi:10.1103/PhysRevD.97.084037
	[arXiv:1711.07431 [hep-th]].
	
	

	\bibitem{Babichev:2017guv} E.~Babichev, C.~Charmousis and A.~Leh\'ebel,
	JCAP \textbf{04} (2017), 027
	doi:10.1088/1475-7516/2017/04/027
	[arXiv:1702.01938 [gr-qc]].

	\bibitem{Sanchez:1977si} N.~G.~Sanchez,
	Phys. Rev. D \textbf{18} (1978), 1030
	doi:10.1103/PhysRevD.18.1030.
	
	
	
	

	\bibitem{Sanchez:1977vz} N.~G.~Sanchez,
	Phys. Rev. D \textbf{18} (1978), 1798
	doi:10.1103/PhysRevD.18.1798

	\bibitem{Crispino:2007zz} L.~C.~B.~Crispino, E.~S.~Oliveira and G.~E.~A.~Matsas,
	Phys. Rev. D \textbf{76} (2007), 107502
	doi:10.1103/PhysRevD.76.107502.
	
	
	

	\bibitem{Crispino:2009ki} L.~C.~B.~Crispino, S.~R.~Dolan and E.~S.~Oliveira,
	Phys. Rev. D \textbf{79} (2009), 064022
	doi:10.1103/PhysRevD.79.064022
	[arXiv:0904.0999 [gr-qc]].
	
	

	\bibitem{Chen:2011jgd} J.~Chen, H.~Liao, Y.~Wang and T.~Chen,
	Eur. Phys. J. C \textbf{73} (2013) no.4, 2395
	doi:10.1140/epjc/s10052-013-2395-9
	[arXiv:1111.0825 [gr-qc]].
	
	
	

	\bibitem{Macedo:2013afa} C.~F.~B.~Macedo, L.~C.~S.~Leite, E.~S.~Oliveira, S.~R.~Dolan and L.~C.~B.~Crispino,
	Phys. Rev. D \textbf{88} (2013) no.6, 064033
	doi:10.1103/PhysRevD.88.064033
	[arXiv:1308.0018 [gr-qc]].
	
	

	\bibitem{Macedo:2014uga} C.~F.~B.~Macedo and L.~C.~B.~Crispino,
	Phys. Rev. D \textbf{90} (2014) no.6, 064001
	doi:10.1103/PhysRevD.90.064001
	[arXiv:1408.1779 [gr-qc]].
	
	
	

	\bibitem{Anacleto:2019tdj} M.~A.~Anacleto, F.~A.~Brito, J.~A.~V.~Campos and E.~Passos,
	Phys. Lett. B \textbf{803} (2020), 135334
	doi:10.1016/j.physletb.2020.135334
	[arXiv:1907.13107 [hep-th]].
	

	\bibitem{Lima:2020auu} H.~C.~D.~Lima, C.~L.~Benone and L.~C.~B.~Crispino,
	Phys. Rev. D \textbf{101} (2020) no.12, 124009
	doi:10.1103/PhysRevD.101.124009
	[arXiv:2006.03967 [gr-qc]].
	

	\bibitem{Anacleto:2020lel} M.~A.~Anacleto, F.~A.~Brito, J.~A.~V.~Campos and E.~Passos,
	Phys. Lett. B \textbf{810} (2020), 135830
	doi:10.1016/j.physletb.2020.135830
	[arXiv:2003.13464 [gr-qc]].
	

	\bibitem{Li:2021epb} Y.~Li and Y.~G.~Miao,
	Phys. Rev. D \textbf{105} (2022) no.4, 044031
	doi:10.1103/PhysRevD.105.044031
	[arXiv:2108.06470 [gr-qc]].

	\bibitem{Li:2022wzi} Q.~Li, C.~Ma, Y.~Zhang, Z.~W.~Lin and P.~F.~Duan,
	Eur. Phys. J. C \textbf{82} (2022) no.7, 658
	doi:10.1140/epjc/s10052-022-10623-3
	[arXiv:2307.04144 [gr-qc]].
	
	
	

	\bibitem{Sun:2023woa} Q.~Sun, Q.~Li, Y.~Zhang and Q.~Q.~Li,
	Mod. Phys. Lett. A \textbf{38} (2023) no.22n23, 2350102
	doi:10.1142/S021773232350102X
	[arXiv:2302.10758 [physics.gen-ph]].
	
	

	\bibitem{Wan:2022vcp} M.~Y.~Wan and C.~Wu,
	Gen. Rel. Grav. \textbf{54} (2022) no.11, 148
	doi:10.1007/s10714-022-03034-y
	[arXiv:2212.01798 [gr-qc]].
	
	
	

	\bibitem{Jung:2004yh} E.~Jung and D.~K.~Park,
	Class. Quant. Grav. \textbf{21} (2004), 3717-3732
	doi:10.1088/0264-9381/21/15/007
	[arXiv:hep-th/0403251 [hep-th]].
	
	

	\bibitem{Benone:2014qaa} C.~L.~Benone, E.~S.~de Oliveira, S.~R.~Dolan and L.~C.~B.~Crispino,
	Phys. Rev. D \textbf{89} (2014) no.10, 104053
	doi:10.1103/PhysRevD.89.104053
	[arXiv:1404.0687 [gr-qc]].
	

	\bibitem{deCesare:2023rmg} M.~de Cesare and R.~Oliveri,
	Phys. Rev. D \textbf{108} (2023) no.4, 044050
	doi:10.1103/PhysRevD.108.044050
	[arXiv:2305.04970 [gr-qc]].
	
	

	\bibitem{Dolan:2006vj} S.~Dolan, C.~Doran and A.~Lasenby,
	Phys. Rev. D \textbf{74} (2006), 064005
	doi:10.1103/PhysRevD.74.064005
	[arXiv:gr-qc/0605031 [gr-qc]].
	

	\bibitem{Sporea:2017zxe} C.~A.~Sporea,
	Chin. Phys. C \textbf{41} (2017) no.12, 123101
	doi:10.1088/1674-1137/41/12/123101
	[arXiv:1707.08374 [gr-qc]].
	
	
	
	
	

	\bibitem{Crispino:2007qw} L.~C.~B.~Crispino, E.~S.~Oliveira, A.~Higuchi and G.~E.~A.~Matsas,
	Phys. Rev. D \textbf{75} (2007), 104012
	doi:10.1103/PhysRevD.75.104012.
	

	\bibitem{Crispino:2008zz} L.~C.~B.~Crispino and E.~S.~Oliveira,
	Phys. Rev. D \textbf{78} (2008), 024011
	doi:10.1103/PhysRevD.78.024011.
	
	

	\bibitem{Crispino:2009xt} L.~C.~B.~Crispino, S.~R.~Dolan and E.~S.~Oliveira,
	Phys. Rev. Lett. \textbf{102} (2009), 231103
	doi:10.1103/PhysRevLett.102.231103
	[arXiv:0905.3339 [gr-qc]].
	

	\bibitem{Leite:2017zyb} L.~C.~S.~Leite, S.~R.~Dolan and L.~C.~B.~Crispino,
	Phys. Lett. B \textbf{774} (2017), 130-134
	doi:10.1016/j.physletb.2017.09.048
	[arXiv:1707.01144 [gr-qc]].
	
	

	\bibitem{Leite:2018mon} L.~C.~S.~Leite, S.~Dolan and L.~Crispino, C.B.,
	Phys. Rev. D \textbf{98} (2018) no.2, 024046
	doi:10.1103/PhysRevD.98.024046
	[arXiv:1805.07840 [gr-qc]].
	

	\bibitem{deOliveira:2019tlk} E.~S.~de Oliveira,
	Eur. Phys. J. Plus \textbf{135} (2020) no.11, 880
	doi:10.1140/epjp/s13360-020-00876-w
	[arXiv:1904.11612 [gr-qc]].
	
	
	
	
	

	\bibitem{Dolan:2008kf} S.~R.~Dolan,
	Class. Quant. Grav. \textbf{25} (2008), 235002
	doi:10.1088/0264-9381/25/23/235002
	[arXiv:0801.3805 [gr-qc]].
	
	

	\bibitem{Crispino:2015gua} L.~C.~B.~Crispino, S.~R.~Dolan, A.~Higuchi and E.~S.~de Oliveira,
	Phys. Rev. D \textbf{92} (2015) no.8, 084056
	doi:10.1103/PhysRevD.92.084056
	[arXiv:1507.03993 [gr-qc]].
	
	
	

	\bibitem{Matzner:1985rjn} R.~A.~Matzner, C.~DeWitte-Morette, B.~Nelson and T.~R.~Zhang,
	Phys. Rev. D \textbf{31} (1985) no.8, 1869
	doi:10.1103/PhysRevD.31.1869.

	\bibitem{Anninos:1992ih} P.~Anninos, C.~DeWitt-Morette, R.~A.~Matzner, P.~Yioutas and T.~R.~Zhang,
	Phys. Rev. D \textbf{46} (1992), 4477-4494
	doi:10.1103/PhysRevD.46.4477.

	\bibitem{Ford:2000uye} K.~W.~Ford and J.~A.~Wheeler,
	Annals Phys. \textbf{281} (2000) no.1-2, 608-635
	doi:10.1016/0003-4916(59)90026-0.

	\bibitem{Misner:1972kx} C.~W.~Misner,
	Phys. Rev. Lett. \textbf{28} (1972), 994-997
	doi:10.1103/PhysRevLett.28.994.

	\bibitem{Brito:2015oca} R.~Brito, V.~Cardoso and P.~Pani,
	Physics,''
	Lect. Notes Phys. \textbf{906} (2015), pp.1-237
	2020,
	ISBN 978-3-319-18999-4, 978-3-319-19000-6, 978-3-030-46621-3, 978-3-030-46622-0
	doi:10.1007/978-3-319-19000-6
	[arXiv:1501.06570 [gr-qc]].
	

	\bibitem{Benone:2015bst} C.~L.~Benone and L.~C.~B.~Crispino,
	Phys. Rev. D \textbf{93} (2016) no.2, 024028
	doi:10.1103/PhysRevD.93.024028
	[arXiv:1511.02634 [gr-qc]].
	

	\bibitem{Benone:2019all} C.~L.~Benone and L.~C.~B.~Crispino,
	Phys. Rev. D \textbf{99} (2019) no.4, 044009
	doi:10.1103/PhysRevD.99.044009
	[arXiv:1901.05592 [gr-qc]].
	

	\bibitem{dePaula:2024xnd} M.~A.~A.~de Paula, L.~C.~S.~Leite, S.~R.~Dolan and L.~C.~B.~Crispino,
	Phys. Rev. D \textbf{109} (2024) no.6, 064053
	doi:10.1103/PhysRevD.109.064053
	[arXiv:2401.01767 [gr-qc]].
	

	\bibitem{Glampedakis:2001cx} K.~Glampedakis and N.~Andersson,
	Class. Quant. Grav. \textbf{18} (2001), 1939-1966
	doi:10.1088/0264-9381/18/10/309
	[arXiv:gr-qc/0102100 [gr-qc]].
	
	

	\bibitem{Feng:2015wvb} X.~H.~Feng, H.~S.~Liu, H.~L\"u and C.~N.~Pope,
	Phys. Rev. D \textbf{93} (2016) no.4, 044030
	doi:10.1103/PhysRevD.93.044030
	[arXiv:1512.02659 [hep-th]].
	

	\bibitem{Wang:2019cuf} C.~Y.~Wang, Y.~F.~Shen and Y.~Xie,
	JCAP \textbf{04} (2019), 022
	doi:10.1088/1475-7516/2019/04/022
	[arXiv:1902.03789 [gr-qc]].
	
	

	\bibitem{Gao:2023mjb} X.~J.~Gao, T.~T.~Sui, X.~X.~Zeng, Y.~S.~An and Y.~P.~Hu,
	Eur. Phys. J. C \textbf{83} (2023), 1052
	doi:10.1140/epjc/s10052-023-12231-1
	[arXiv:2311.11780 [gr-qc]].
	

	\bibitem{Unruh:1976fm} W.~G.~Unruh,
	Phys. Rev. D \textbf{14} (1976), 3251-3259
	doi:10.1103/PhysRevD.14.3251
	
	%

	\bibitem{Abramowitz} M.~Abramowitz, I. A. Stegun and D.~Miller, \textit{Handbook of mathematical functions with formulas, graphs and mathematical tables} (Dover Publications, New York, 1965), 9th ed.

	\bibitem{Zhou:2022dze} S.~Zhou, M.~Chen and J.~Jia,
	Eur. Phys. J. C \textbf{83} (2023) no.9, 883
	doi:10.1140/epjc/s10052-023-12047-z
	[arXiv:2203.05415 [gr-qc]].
	
	
	%

	\bibitem{Wald2010} Wald, Robert M. General relativity. University of Chicago press, 2010.

	\bibitem{Collins:1973xf} P.~A.~Collins, R.~Delbourgo and R.~M.~Williams,
	J. Phys. A \textbf{6} (1973), 161-169
	doi:10.1088/0305-4470/6/2/007.
	
	%

	\bibitem{Xu:2021rld} X.~Xu, T.~Jiang and J.~Jia,
	JCAP \textbf{08} (2021), 022
	doi:10.1088/1475-7516/2021/08/022
	[arXiv:2105.12413 [gr-qc]].

	\bibitem{Pang:2018jpm} X.~Pang and J.~Jia,
	Class. Quant. Grav. \textbf{36} (2019) no.6, 065012
	doi:10.1088/1361-6382/ab0512
	[arXiv:1806.04719 [gr-qc]].
	
	
	%

	\bibitem{Zhang:1984vt} T.~R.~Zhang and C.~DeWitt-Morette,
	Phys. Rev. Lett. \textbf{52} (1984), 2313-2316
	doi:10.1103/PhysRevLett.52.2313.
	
	

	\bibitem{Dolan:2009zza} S.~R.~Dolan, E.~S.~Oliveira and L.~C.~B.~Crispino,
	Phys. Rev. D \textbf{79} (2009), 064014
	doi:10.1103/PhysRevD.79.064014
	[arXiv:0904.0010 [gr-qc]].

	\bibitem{Yennie:1954zz} D.~R.~Yennie, D.~G.~Ravenhall and R.~N.~Wilson,
	Phys. Rev. \textbf{95} (1954), 500-512
	doi:10.1103/PhysRev.95.500.

	\bibitem{Cotaescu:2014jca} I.~I.~Cotaescu, C.~Crucean and C.~A.~Sporea,
	Eur. Phys. J. C \textbf{76} (2016) no.3, 102
	doi:10.1140/epjc/s10052-016-3936-9
	[arXiv:1409.7201 [gr-qc]].


\end{thebibliography}
\end{document}